\documentclass[iop,numberedappendix]{emulateapj}

\usepackage{amsmath}
\usepackage{color}
\usepackage{hyperref}

\newcommand{\pext}{p_{\rm ext}}
\newcommand{\p}{\partial}
\newcommand{\kx}{k_x}
\newcommand{\Msun}{M_\odot}
\newcommand{\poo}{p_{00}}
\newcommand{\rhoo}{\rho_{00}}
\newcommand{\sech}{\mathrm{sech}}
\newcommand{\Qeff}{Q_{\rm eff}}
\newcommand{\Qc}{Q_{{\rm eff},c}}
\newcommand\simgt{\lower.5ex\hbox{$\; \buildrel > \over \sim \;$}}
\newcommand\simlt{\lower.5ex\hbox{$\; \buildrel < \over \sim \;$}}

\long\def\symbolfootnote[#1]#2{\begingroup%
\def\thefootnote{\fnsymbol{footnote}}\footnote[#1]{#2}\endgroup}

\shorttitle{Gravitational Instability of Disks} %%
\shortauthors{Kim et al.}

\begin{document}

\title{Gravitational Instability of Rotating, Pressure-Confined,
Polytropic Gas Disks With Vertical Stratification}

\author{Jeong-Gyu Kim\altaffilmark{1,2},
Woong-Tae Kim\altaffilmark{1,2},
Young Min Seo\altaffilmark{2,3},
\& Seung Soo Hong\altaffilmark{2,4}}

\affil{$^1$Center for the Exploration of the Origin of the Universe
  (CEOU), Astronomy Program, Department of Physics \& Astronomy,\\ Seoul
  National University, Seoul 151-742, Republic of Korea}
\affil{$^2$FPRD, Department of Physics \& Astronomy, Seoul National
  University, Seoul 151-742, Republic of Korea} %%
\affil{$^3$Department of Astronomy and Steward Observatory, University
  of Arizona, Tucson, AZ 85721, USA} %%
\affil{$^4$National Youth Space Center, Goheung, Jeolanamdo, 548-951,
  Republic of Korea}
\email{jgkim@astro.snu.ac.kr}
\email{wkim@astro.snu.ac.kr}
\email{seo3919@email.arizona.edu}
\email{sshong@astro.snu.ac.kr}
\slugcomment{Accepted for publication in the \apj}
%\slugcomment{\sc{Accepted to ApJ:} October 22, 2012}
%\altaffiltext{2}{Corresponding author: wkim@astro.snu.ac.kr}

\begin{abstract}
We investigate gravitational instability (GI) of rotating,
vertically-stratified, pressure-confined, polytropic gas disks using
linear stability analysis as well as analytic approximations. The
disks are initially in vertical hydrostatic equilibrium and bounded by
a constant external pressure. We find that GI of a pressure-confined
disk is in general a mixed mode of the conventional Jeans and
distortional instabilities, and is thus an unstable version of
acoustic-surface-gravity waves. The Jeans mode dominates in weakly
confined disks or disks with rigid boundaries.  When the disk has free
boundaries and is strongly pressure-confined, on the other hand, the
mixed GI is dominated by the distortional mode that is surface-gravity
waves driven unstable under own gravity and thus incompressible. We
demonstrate that the Jeans mode is gravity-modified acoustic waves
rather than inertial waves and that inertial waves are almost
unaffected by self-gravity. We derive an analytic expression for the
effective sound speed $c_{\rm eff}$ of acoustic-surface-gravity
waves. We also find expressions for the gravity reduction factors
relative to a razor-thin counterpart, appropriate for the Jeans and
distortional modes. The usual razor-thin dispersion relation after
correcting for $c_{\rm eff}$ and the reduction factors closely matches
the numerical results obtained by solving a full set of linearized
equations. The effective sound speed generalizes the Toomre stability
parameter of the Jeans mode to allow for the mixed GI of
vertically-stratified, pressure-confined disks.
\end{abstract}


\section{Introduction}\label{sec:intro}

Gravitational Instability (GI) plays a crucial role in structure
formation in various astronomical situations ranging from growth of
primordial density fluctuations to star formation in disk galaxies
(e.g., \citealt{Zeldovich70, McKee+07}). GI in flattened systems is
of particular importance in formation of giant clouds in disk
galaxies (e.g., \citealt{Goldreich+65a, Elmegreen87, Kim+03}), gas
giant planets in protoplanetary disks (e.g., \citealt{Boss97,
Durisen+07}), and bound clumps resulting from fragmentation of
shocked layers in star-forming regions (e.g.,
\citealt{Elmegreen+77,Elmegreen98}).

While real astrophysical disks are vertically stratified, most
analytic studies often neglect the vertical degree of freedom and
adopt an isothermal equation of state (EOS) for simplicity.  In a
rotating, infinitesimally-thin disk with surface density $\Sigma_0$
and sound speed $c_s$, the local dispersion relation for
axisymmetric waves with frequency $\omega$ and wavenumber $k$ is
given by
\begin{equation} \label{eq:disp_razor}
\omega^2 = c_s^2k^2 - 2\pi G\Sigma_0 k + \kappa_0^2,
\end{equation}
where $\kappa_0$ is the epicycle frequency and $G$ is the
gravitational constant (e.g., \citealt{Goldreich+78, Binney+08}). If
the disk is not rotating, sound waves with $k<k_{\rm J}=2\pi G
\Sigma_0/c_s^2$ would become gravitationally unstable. The Coriolis
force arising from disk rotation stabilizes long-wavelength
perturbations, making the disk Jeans-stable if the Toomre stability
parameter satisfies
\begin{equation}\label{eq:TooQ}
Q_T \equiv \frac{c_s\kappa_0}{\pi G \Sigma_0} > Q_{T,c}=1,
\end{equation}
for any $k$ (e.g., \citealt{Toomre64}).  When $Q_{T,c}<1$, on the
other hand, thermal pressure and rotation are unable to stop the
collapse of over-dense regions.

The thin-disk approximation would be valid as long as vertical
motions are unimportant and the scale of interest is much longer
than the disk scale height.  Nevertheless, it has been well known
that finite disk thickness makes some quantitative changes to the
characteristic wavelengths and critical $Q_T$ values.  For instance,
\citet{Ledoux51} showed that waves in self-gravitating,
non-rotating, isothermal disks with scale height $H_0=c_s^2/(\pi G
\Sigma_0)$ become unstable if $kH_0 < 1$, which can be compared with
the unstable condition $kH_0<2$ of the razor-thin counterpart (see
also \citealt{Simon65}).  For rotating disks, \citet{Goldreich+65a}
considered the effect of the finite disk thickness and found that
the stability condition changes to $Q_{T,c}= 0.676$. This decrease
of the critical $Q_T$ value is due to the dilution of self-gravity
at the disk midplane in a vertically stratified disk, which can
approximately be treated by multiplying the gravity reduction factor
$\mathcal{F}=1/(1+kH_0)$ to the second term of the right-hand-side
of equation \eqref{eq:disp_razor} \citep[e.g.,][]{Elmegreen87,
Kim+02,Kim+07}.

While the results of GI in isothermal disks are informative, there
are several issues of GI that need clarification in more general
situations. First of all, there are many classes of gaseous disks
that do not have constant temperature in the vertical direction.
Examples include optically thick regions of planet-forming
protoplanetary disks (e.g., \citealt{Bell+97, Dalessio+98,
  Boley+07}) and accretion disks around compact objects
(e.g., \citealt{LaDous94,Lubow+98}). These disks have often been
modeled using a polytropic EOS \citep[e.g.,][]{Lubow+98, Nelson+98,
Mamatsashvili+10}. While behavior of various waves in polytropic
disks have been studied extensively (e.g., \citealt{Lin+90,
Korycansky+95, Lubow+98, Ogilvie+99}), GI of such disks has not been
explored in detail. \citet{Goldreich+65a} calculated a stability
criterion of a uniformly-rotating polytropic disk, but was limited
to the case with the adiabatic index $\gamma=2$. \citet{Larson85}
studied GI of polytropic disks with arbitrary $\gamma$ by using
approximate scaling relations instead of solving perturbation
equations accurately. Recently, \citet{Mamatsashvili+10} considered
vertically-stratified polytropic disks and argued that inertial
modes rather than acoustic modes become unstable to self-gravitating
perturbations. In this work, we shall clarify using both analytic
and numerical approaches that it is the acoustic modes rather than
the inertial modes that become unstable.

Second, since the sound speed varies with height in polytropic
disks, it is questionable what kind of average is the most
appropriate for $c_s$ if one wants to use the stability condition
\eqref{eq:TooQ}.  Several numerical studies used the
vertically-averaged sound speed \citep{Mejia+05, Boley+06} or simply
midplane value \citep{Rice+03,Rice+05} to evaluate stability of
protoplanetary disks, but the usage of these has yet to be
justified. \citet{Mamatsashvili+10} attempted to resolve this
ambiguity by introducing $Q_M=\Omega^2/(4\pi G\rhoo)$ as a
three-dimensional stability parameter, where $\Omega$ is the local
orbital angular velocity and $\rhoo$ is the midplane density, but
the connection between $Q_M$ and $Q_T$ is uncertain since the former
does not depend on $c_s$ explicitly. In this work, we define the
effective sound speed that well represents the modal behavior of
acoustic waves in a thermally stratified disk and show that this
naturally relates $Q_T$ to $Q_M$.

The third issue involves disk truncation by external pressure.
Galactic disks are confined by ram pressure of infalling gas (e.g.,
\citealt{Dubois+08}) or hot halo gas (e.g., \citealt{Goldreich+65a,
Lee+07}). Thin shells produced by supernovae, stellar winds, or
expanding \ion{H}{2} regions are usually bounded by shocks
\citep[e.g.,][]{Deharveng+05, Churchwell+06, Churchwell+07}.
\citet{Elmegreen+78} showed that pressure-confined disks become
unstable at scales smaller than the Jeans wavelength
$\lambda_{\rm{J}} =c_s^2/(\pi G \Sigma_0)$ of unbounded disks. The
mass of fragments produced by such GI is less than the critical
Bonnor-Ebert mass, and thus may not necessarily experience
gravitational runaway. When the confining pressure is very strong,
\citet{Lubow+93} showed that the GI becomes essentially the same as
that of an incompressible disk, which is in stark contrast to the
conventional Jeans instability of compressional disks.
\citet{Umekawa+99} and \citet{Wunsch+10} confirmed such predictions
by using numerical simulations. \citet{Boyd+05} suggested the GI of
strongly confined disks as a potential candidate for forming
``free-floating'' planetary-mass objects (having mass as low as
$\sim0.003\Msun$) detected in young star clusters
\citep[e.g.,][]{Zapatero+02}. More elaborate theoretical and
numerical models including the effects of shell expansion and
deceleration \citep[e.g.,][]{Elmegreen89, Iwasaki+11, Dale+11},
thermal and chemical processes \citep{Hosokawa+06}, boundary
conditions \citep[e.g.,][]{Elmegreen89, Usami+95, Dale+09}, magnetic
fields \citep{Nagai+98}, etc.\ examined consequences of GI occurring
under high external pressure.

Despite these efforts, however, the physical mechanism behind the GI
of strongly pressure-confined disks still remains controversial. For
example, \citet{Lubow+93} argued that the instability is due to a
neutral mode that exists because external pressure can hold the
layer to any distorted shape, while \citet{Elmegreen89} and
\citet{Umekawa+99} claimed that the distortion of the surfaces
exerts a pinching force that causes the distorted disk to collapse.
On the other hand, \citet{Wunsch+10} and \citet{Dale+11} considered
the collapse of an oblate spheroid with uniform density pressurized
by an ambient medium (see also \citealt{Boyd+05}), and termed the
enhanced GI ``pressure-assisted gravitational instability'',
although the physical processes responsible for it were not
identified.  Very recently, \citet{Iwasaki+11} studied GI of
pressure-confined shells around expanding \ion{H}{2} regions taking
into account boundary effects. They suggested from the vertical
behaviors of eigenfunctions that the restoring force of the unstable
mode comes from surface-gravity waves. We use both numerical and
analytic approaches to confirm that it is indeed surface-gravity
waves that become unstable in strongly confined disks.

In this paper, we investigate GI of rotating, vertically-stratified,
pressure-confined gas disks.  We adopt simple polytropic, rotating
models of \cite{Goldreich+65a} and extend them to pressure-confined
situations. This work also extends the case of non-rotating,
isothermal models of \citet{Elmegreen+78} to rotating, polytropic
disks. Our objectives are four-fold. First, we wish to distinguish
conventional Jeans instability from the GI arising from the surface
distortion of strongly pressure-confined disks, and give a clear
physical explanation for the latter that has previously been
confusing. Second, we provide analytic expressions for the effective
sound speed $c_{\rm eff}$ that best represents the modal behavior of
sound waves in polytropic disks.  Third, we find the gravity
reduction factors caused by finite disk thickness for the Jeans and
distortional modes of GI.  We show that the numerical dispersion
relations found by solving a full set of perturbation equations are
well matched by equation \eqref{eq:disp_razor} provided that $c_s$
is changed to $c_{\rm eff}$ and that the appropriate reduction
factor is considered in the gravity term. Finally, we generalize the
Toomre stability parameter of razor-thin disks to
vertically-stratified, pressure-bounded disks. While our disk models
are simple with a polytropic EOS and without considering external
gravity, some qualitative properties of GI clarified in this paper
can hold true in more general situations such as disks with a
barotropic EOS and/or with external gravity.

The remainder of this paper is organized as follows.  In Section
\ref{sec:basic}, we introduce basic equations, construct an
equilibrium model, and present linearized perturbation equations. In
Section \ref{sec:fun}, we derive expressions for the effective sound
speed $c_{\rm eff}$ that describe the phase speeds of fundamental
modes in non-self-gravitating, non-rotating disks.  In Section
\ref{sec:gi}, we present the gravity reduction factors
$\mathcal{F}$, and show that an approximate dispersion relation
using $c_{\rm eff}$ and $\mathcal{F}$ is in excellent agreement with
the numerical results obtained by solving the perturbation
equations. For rotating disks, we define a stability parameter using
$c_{\rm eff}$ that is applicable also to the distortional mode. In
Section \ref{sec:discuss}, we summarize and discuss our results.

\section{Formulations}\label{sec:basic}

We consider a rotating, self-gravitating, gaseous disk in vertical
hydrostatic equilibrium and analyze its stability to
small-amplitude, axisymmetric perturbations.  The disk is infinitely
extended in the horizontal direction, but truncated at a finite
height by a tenuous external medium with pressure $\pext$. For local
modes in the horizontal direction, it is advantageous to consider a
local Cartesian reference frame whose center lies at a radius $R_0$
and orbits the disk center with a fixed angular velocity
$\Omega_0=\Omega(R_0)$.  In this local frame, radial, azimuthal, and
vertical coordinates are represented by $x=R-R_0$,
$y=R_0(\phi-\Omega_0 t)$, and $z$, respectively, and the terms
arising from curvature effect are ignored (e.g.,
\citealt{Goldreich+65b,Julian+66,Kim+02}). Assuming that the shear
rate $ q \equiv - d \ln \Omega / d \ln R |_{R_0}$ is uniform, the
equilibrium background velocity field is given by $\mathbf{u}_0 = -q
\Omega_0 x \mathbf{e}_y$ and the local epicyclic frequency is
$\kappa_0^2 = (4-2q)\Omega_0^2$. The equations of ideal
hydrodynamics expanded in the local frame are
\begin{equation}\label{eq:con}
\frac{\p \rho}{\p t} + \nabla \cdot (\rho \mathbf{u}) = 0,
\end{equation}
\begin{equation}\label{eq:mom}
\frac{\p \mathbf{u}}{\p t} + \mathbf{u}\cdot\nabla\mathbf{u} = -
\frac{1}{\rho}\nabla p -\nabla \psi + 2q \Omega_0^2 x \mathbf{e}_x - 2
\mathbf{\Omega}_0 \times \mathbf{u},
\end{equation}
\begin{equation}\label{eq:Poi}
\nabla^2 \psi = 4\pi G \rho,
\end{equation}
where $\rho$, $\mathbf{u}$, $p$, and $\psi$ denote mass density,
velocity, thermal pressure, and self-gravitational potential of the
gas, respectively. To close the set of equations
\eqref{eq:con}-\eqref{eq:Poi}, we assume that the gas follows a
polytropic EOS
\begin{equation}
p = K\rho^{\gamma},
\end{equation}
where $K$ is an arbitrary constant and $\gamma$ is the polytropic
exponent. If $\gamma$ is different from an adiabatic exponent of
disturbances, the system under consideration exhibits complicated
modal behaviors associated with convective motions. However,
\citet{Mamatsashvili+10} showed that the convective modes do not
affect the properties of GI much. In this work, we thus simply take
$\gamma$ equal to the adiabatic exponent.

\subsection{Initial Equilibria}

We first explore equilibrium profiles of density and temperature in
purely self-gravitating, polytropic disks confined by an external
medium. The pressure confinement model we adopt is a straightforward
extension of isothermal disks studied by \citet{Elmegreen+78} to
polytropic cases. For hydrostatic equilibrium in the vertical
direction, equations (\ref{eq:mom}) and (\ref{eq:Poi}) require
\begin{equation}\label{eq:HSE1}
\frac{1}{\rho_0}\frac{dp_0}{d z} = -\frac{d\psi_0}{d z},
\end{equation}
\begin{equation} \label{eq:HSE2}
\frac{d^2}{d z^2}\psi_{0} = 4\pi G\rho_0,
\end{equation}
where the subscript ``0'' denotes the unperturbed state.

Following \citet{Elmegreen+78}, it is convenient to
convert the vertical coordinate $z$ to the dimensionless variable $\mu$
defined by
\begin{equation}\label{eq:mu}
\mu \equiv \frac{1}{4\pi G\rhoo H_0}\frac{d\psi_0}{dz},
\end{equation}
where $\rhoo=\rho_0(0)$ and $H_0$ is the disk scale height
\begin{equation}\label{eq:H0}
H_0^2 \equiv \frac{K}{2\pi G\rhoo^{2-\gamma}}.
\end{equation}
In Appendix \ref{app:ini}, we show that equations (\ref{eq:HSE1})
and (\ref{eq:HSE2}) subject to the symmetry condition
${d\rho_0}/{dz}=0$ at $z=0$ yield the solution
\begin{equation}\label{eq:density}
\rho_0 = \rhoo(1-\mu^2)^{1/\gamma},
\end{equation}
(see also \citealt{Harrison+72}).
The corresponding local speed of sound is
\begin{equation}
c_{s}^2 = \gamma p_0 / \rho_0 = c_{s0}^2(1-\mu^2)^{1-1/\gamma},
\end{equation}
where $c_{s0} \equiv (\gamma \poo/\rhoo)^{1/2} =
(\gamma K\rhoo^{\gamma-1})^{1/2}$ is the midplane sound speed.

%% Figure1: Vertical equilibrium structure
\begin{figure}
\epsscale{1.15}\plotone{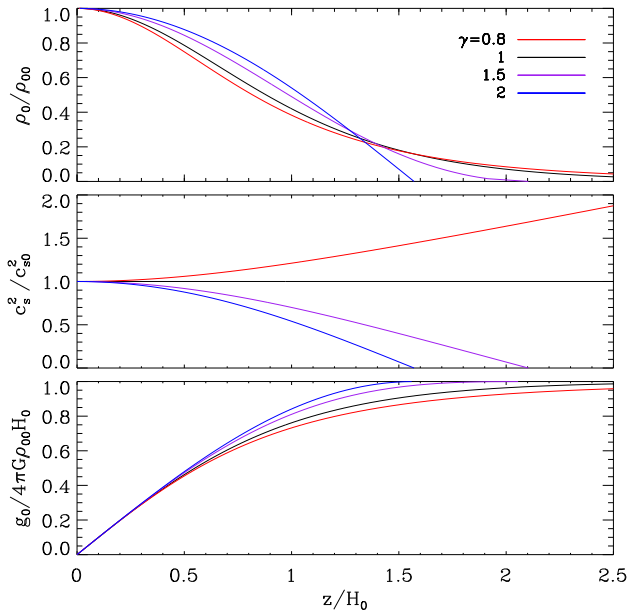}
\caption{ Distributions of density $\rho_0$ (top), local sound speed
  $c_s$ (middle), and vertical gravity $g_0=d\psi_0/dz$ (bottom) in
  hydrostatic equilibrium of self-gravitating polytropic disks with
  exponent $\gamma$. A stiffer equation of state (larger $\gamma$)
  leads to a flatter density profile near the midplane. Even without
  external pressure, disks with $\gamma > 1$ are truncated at a finite
  height (eq. [\ref{eq:trun}]).  For $\gamma < 1$, $c_s$ is an
  increasing function of $z$.}\label{fig:1}
\end{figure}

Figure \ref{fig:1} plots equilibrium structures of self-gravitating
polytropic disks with differing $\gamma$. For a softer EOS (smaller
$\gamma$), mass density falls off more rapidly from the midplane to
provide the pressure support against the gravity $\sim 4\pi G\rhoo
z$ at $|z|/H_0 \lesssim 1$. This in turn makes the gravity smaller
at high-altitude regions, causing the density to decrease slower
with height at $|z|/H_0 > 1$. As $\gamma$ increases, therefore, the
density becomes more concentrated toward the midplane. For
$\gamma>1$, equation \eqref{eq:z_over_H} implies that the density
becomes automatically zero at finite height
\begin{equation}\label{eq:trun}
\frac{|z_{\rm cut}|}{H_0} = \frac{\sqrt{\pi}}{2}\frac{\Gamma(1 -
1/\gamma)}{\Gamma(3/2 - 1/\gamma)}\,,
\end{equation}
where $\Gamma$ is the Gamma function, indicating that disks with
$\gamma>1$ are truncated even without external pressure. The sound
speed in a disk with $\gamma>1$ decreases monotonically with
increasing $z$. Note that $|z_{\rm cut}| =\infty$ for $\gamma\leq1$.

Now we allow for the truncation of a disk at height $z=\pm a$ due to
an external pressure $\pext$. The condition of pressure equilibrium
at the boundaries requires
\begin{equation}\label{eq:A}
A \equiv \mu(z=\pm a) = \left(1+\frac{2\pext}{\pi G \Sigma_0^2}
\right)^{-1/2},
\end{equation}
(e.g., \citealt{Elmegreen+78}). Using equation \eqref{eq:zH0}, the
column density of the bounded layer is given by
\begin{equation}
\Sigma_0 = \int_{-a}^{+a}\rho_0\,d z = 2\rhoo AH_0 = 2\rhoo H\,,
\end{equation}
where $H \equiv AH_0$ is the effective half-thickness of the layer.
The relative density drop at the boundaries is $\rho_{0a}/\rhoo = (1
- A^2)^{1/\gamma}$ where $\rho_{0a} = \rho_0(z = a)$. Note that
$A=1$ when $a=|z_{\rm cut}|$, corresponding to no pressure
truncation. As $\pext$ increases, the vertical sound crossing time
($\sim H/c_{s0}$) becomes smaller than the internal free-fall time
($\sim (2\pi G\rhoo)^{-1/2}$). In the limit of $A \rightarrow 0$,
the disk takes uniform density and $H \rightarrow a$ regardless of
$\gamma$: in this case, self-gravity is so weak that the initial
equilibrium is maintained by balance between $\pext$ and $\poo$.

\subsection{Perturbation Equations}

We apply small-amplitude perturbations to the initial equilibrium
configurations described above. Equations
\eqref{eq:con}--\eqref{eq:Poi} are linearized to
\begin{equation} \label{eq:per_cont}
\frac{\p \rho_1}{\p t} + \nabla \cdot (\rho_0 \mathbf{u}_1) = 0\,,
\end{equation}
\begin{equation} \label{eq:per1}
\frac{\p \mathbf{u}_1}{\p t} = \frac{\rho_1}{\rho_0^2}
\nabla p_0 - \frac{1}{\rho_0}\nabla p_1 - \nabla \psi_1
+ q\Omega_0u_{1x}\mathbf{e}_y - 2 \mathbf{\Omega}_0 \times \mathbf{u}_1\,,
\end{equation}
\begin{equation} \label{eq:per3}
\nabla^2 \psi_1 = 4\pi G\rho_1\,,
\end{equation}
where the subscript ``1'' indicates the perturbed quantities.

We further assume that perturbations are axisymmetric and of plane-wave type
\begin{equation}\label{eq:Fou}
q_1(x,z,t) = q_1(z) \exp (i\omega t + i\kx x)\,,
\end{equation}
where $q_1$ refers to any perturbed variable with frequency $\omega$
and radial wavenumber $\kx$. Then, equations
\eqref{eq:per_cont}--\eqref{eq:per3} can be reduced to four
first-order ordinary differential equations
\begin{eqnarray}
\frac{d\xi_{1z}}{dz} &=& \frac{\kx^2}{\omega^2-\kappa_0^2}(h_1 +
\psi_1)
- \frac{1}{c_s^2}(h_1 - g_0\xi_{1z})\,, \label{eq:per_a} \\
\frac{dh_1}{dz} &=& \omega^2\xi_{1z} - \psi_1^{\prime}\,, \label{eq:per_b}\\
\frac{d\psi_1}{dz} &=& \psi_1^{\prime}\,, \label{eq:per_c} \\
\frac{d\psi_1^{\prime}}{dz}&=& \kx^2\psi_1 + \frac{4\pi
G\rho_0}{c_s^2}h_1\,, \label{eq:per_d}
\end{eqnarray}
where $\xi_{1z}$ is the vertical Lagrangian displacement
defined through $u_{1z}=\p \xi_{1z}/\p t$,
$h_1 = p_1/\rho_0$ is the perturbed enthalpy,
and $\psi_1^{\prime} = d\psi_1/dz$.
We take $\xi_{1z}$, $h_1$, $\psi_1$, and $\psi^{\prime}_1$ as
four independent variables.

The set of the perturbation equations
(\ref{eq:per_a})--(\ref{eq:per_d}) are to be integrated numerically
along the vertical direction to yield a dispersion relation
$\omega=\omega(\kx)$ that satisfies the boundary conditions (BCs) as
well as the constraints of even-symmetry modes.  To separate the
distortional modes of GI from the conventional Jeans modes, we
consider two distinct conditions at the boundaries: the rigid
boundary that allows only the Jeans modes and the free boundary in
which Jeans and distortional modes coexist.  Appendix \ref{app:bc}
describes the BCs we adopt. Appendix \ref{app:num} presents the
numerical method we follow to obtain full dispersion relations,
which will be compared with approximate dispersion relations in
Sections \ref{sec:fun} and \ref{sec:gi}.

For future purposes, we note that the total perturbed density
$\rho_1$ is a superposition of the perturbed density $\rho_{1,i}$
inside the disk due to wave motions and the perturbed density
$\rho_{1,s}$ at $z=\pm a$ due to the surface distortion:
\begin{equation}
\rho_1 = \rho_{1,i} + \rho_{1,s},
\end{equation}
where
\begin{equation}\label{eq:sigma1s}
\rho_{1,s} = \rho_{0a}\xi_{1z}\bigl|_{z=a}\delta (|z| - a),
\end{equation}
for even-symmetry modes. Here, $\delta$ is the Dirac delta function.
The corresponding perturbations in the surface density can be
written as $\Sigma_1 = \Sigma_{1,i} + \Sigma_{1,s}$. We define
\begin{equation}\label{eq:w}
w \equiv \frac{\Sigma_{1,i}}{\Sigma_{1,i} + \Sigma_{1,s}},
\end{equation}
as the fraction of the perturbed surface density inside the disk
relative to the total perturbed surface density. We will show that
$w$ defined in equation \eqref{eq:w} is a key parameter that
controls the relative importance of the Jeans modes to the
distortional modes.

\subsection{Classification of Local Modes}\label{sec:mode}

Ignoring waves arising from surface distortion, the perturbation
equations (\ref{eq:per_a})--(\ref{eq:per_d}) give rise to two
principal types of waves: inertial waves and acoustic waves, both
modified by self-gravity. The former is characterized by epicycle
motions, while the latter is based on gas compressibility.
\citet{Mamatsashvili+10} argued that inertial modes, rather than
acoustic modes, are strongly influenced by self-gravity to become
unstable. In Appendix \ref{app:WKB}, we classify two types of waves
using the local dispersion relations and compare them with the
numerical results. We directly demonstrate that while the inertial
modes are not much affected by self-gravity and thus remain stable,
it is the fundamental acoustic modes that can become unstable in the
presence of self-gravity. This makes sense since both acoustic waves
and self-gravity rely on density perturbations, while inertial waves
are incompressible in nature (e.g., \citealt{Latter+09}).

Before exploring numerical dispersion relations, some physical
insight on the fundamental acoustic waves in disks can be gleaned by
integrating equation \eqref{eq:per_cont} and the horizontal
component of equation \eqref{eq:per1} multiplied by $\rho_0$ over
$z$. With the help of equation \eqref{eq:Fou}, we obtain
\begin{equation}\label{eq:omegasq_int}
\omega^2 = \kx^2\Sigma_0\frac{\widehat{h}_1
+\widehat{\psi}_1}{\Sigma_1} + \kappa_0^2,
\end{equation}
where the hat indicates the density-weighted vertical averages
(i.e., $\widehat{q}_1 = \int \rho_0\,q_1\,dz/\Sigma_0$).  Because
$\widehat{h}_1$, $\widehat{\psi}_1$, and $\Sigma_1$ are interrelated
with each other and generally depend on $\kx$ and $\omega$, one
needs to solve equations \eqref{eq:per_a}--\eqref{eq:per_d}
numerically to obtain full dispersion relations. Nevertheless,
equation \eqref{eq:omegasq_int} implies that $\omega^2$ consists of
three terms, each responsible for the effect of pressure (including
thermal pressure and surface distortion), self-gravity, and the
Coriolis force, as in equation \eqref{eq:disp_razor}. This opens a
possibility that by finding solutions of $\widehat{h}_1$ and
$\widehat{\psi}_1$ independent of $\omega$ in some suitable limits,
one can obtain an approximate dispersion relation that matches the
numerical results closely. We will focus on this in the remainder of
this paper.

\section{Non-self-gravitating Fundamental Modes}\label{sec:fun}

As mentioned before, the sound speed of polytropic disks varies with
the vertical height, so that it is interesting to find an effective
sound speed that best represents modal behavior of waves propagating
throughout the disk.  Also of interest is the effect of surface
distortion on sound waves in the case of the free boundaries. To
address these issues, we in this section limit ourselves to the
fundamental acoustic modes in non-self-gravitating disks. For
simplicity, we ignore the effect of disk rotation, which merely adds
$\kappa_0^2$ to $\omega^2$ (see eq.\ [\ref{eq:omegasq_int}]).

\subsection{Rigid Boundary}

In non-self-gravitating, non-rotating disks, equation
\eqref{eq:omegasq_int} is reduced to
\begin{equation}\label{eq:disp_wav}
\omega^2 = \frac{\Sigma_0\widehat{h}_1}{\Sigma_1}\kx^2 =
\frac{\int\,\rho_0h_1dz}{\Sigma_{1,i} + \Sigma_{1,s}}\kx^2 =
c_{\rm{eff}}^2\kx^2,
\end{equation}
where $c_{\rm{eff}}$ is the effective sound speed. When the
boundaries are rigid or $A=1$, $\Sigma_{1,s} = 0$ and
\begin{equation} \label{eq:c_eff_def}
c_{\rm{eff}}^2 \equiv \frac{\int\rho_{1}c_s^2\,dz}{\int\rho_{1}\,dz}
=
\frac{\int_{-a}^{a}\rho_0h_1\,dz}{\int_{-a}^{a}\rho_0h_1/c_s^2\,dz}.
\end{equation}
Equation \eqref{eq:c_eff_def} shows that excited waves are pure
acoustic modes ($p$ modes) whose restoring force is the thermal
pressure.  Obviously, $c_{\rm{eff}}= c_{s0}$ in an isothermal disk.
Note that we did not make any approximation in deriving equations
\eqref{eq:disp_wav} and \eqref{eq:c_eff_def} except for ignoring the
effects of self-gravity and rotation.

Now consider the fundamental mode in the long-wavelength limit.
Appendix \ref{app:non-self} shows that the fundamental acoustic
waves have a simple solution $h_1(z) = \mathrm{constant}$ in the
limit of $\kx \rightarrow 0$. Equation \eqref{eq:c_eff_def} is then
simplified to
\begin{equation}\label{eq:rigid_approx}
c_{\rm{eff}}^2=\widetilde{c_s}^2 \equiv
\frac{\int_{-a}^{a}\rho_0\,dz}{\int_{-a}^{a}\rho_0/c_s^2\,dz},
\;\;\; \mbox{for rigid BC},
\end{equation}
indicating that the effective sound speed corresponds to the
density-weighted, harmonic mean of the local sound speeds.
For polytropic disks, the ratio of the effective sound speed
to the midplane sound speed is given analytically by
\begin{equation} \label{eq:alpha}
\alpha \equiv \widetilde{c_s}^2/c_{s0}^2%\bigl|_{k \rightarrow 0}
= {{}_{2}F_{1}(\case{1}{2},\,1-\case{1}{\gamma};\,\case{3}{2};\,A^2)}^{-1},
\end{equation}
where ${}_{2}F_{1}$ denotes the Gaussian hypergeometric function.
Disks with $\gamma > 1$ have $\alpha < 1$ since the temperature
decreases monotonically with height. When there is no external
confining medium ($A=1$), $\alpha$ varies smoothly from $3/2$ to
$2/\pi$ for $0.5 < \gamma < 2$. Larger external pressure makes
$\alpha$ closer to unity.

%% Figure2: Sound waves under the rigid BC
\begin{figure}
\epsscale{1.2} \plotone{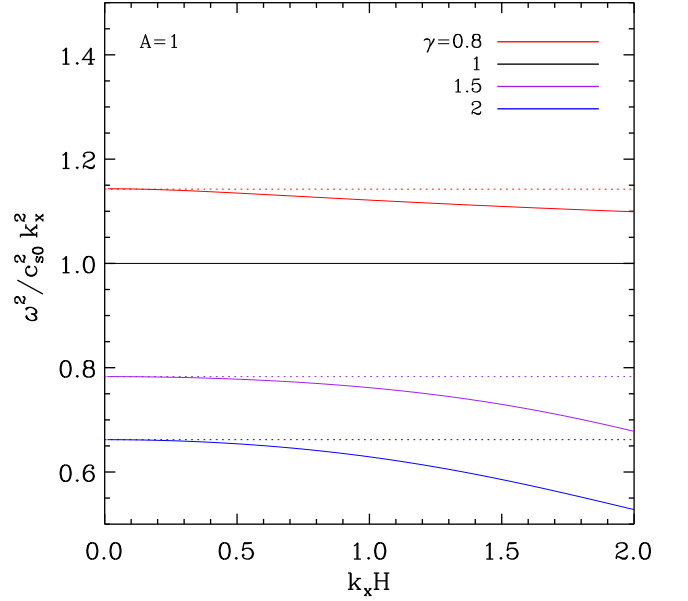}
\caption{Dispersion
relations of the fundamental acoustic modes in
  non-self-gravitating, non-rotating, vertically-stratified,
  polytropic disks with $A=1$ and $\gamma = 0.8, 1, 1.5, 2$. The frequency
  $\omega$ in the ordinate is normalized by $c_{s0}\kx$. Horizontal
  dotted lines mark $c_{\rm eff}^2/c_{s0}^2\,(\equiv \alpha)$
  (eq.\ [\ref{eq:alpha}]) in the long wavelength limit.
}\label{fig:acoustic}
\end{figure}

To check the applicability of equation \eqref{eq:rigid_approx} for
waves with finite wavelengths, we obtain numerical dispersion
relations of pure acoustic waves in non-self-gravitating,
non-rotating disks for $A=1$ and $\gamma =0.8,\,1,\,1.2,$ and $2$,
by following the procedures given in Appendix \ref{app:num}. Figure
\ref{fig:acoustic} plots the resulting numerical dispersion
relations using solid lines, which can be compared with the
approximate dispersion relation \eqref{eq:disp_wav} together with
equation \eqref{eq:rigid_approx} shown as dotted lines.  When
$\gamma = 1$, the dispersion relation is simply $\omega^2 =
c_{s0}^2\kx^2$. When $\gamma \ne 1$, the numerical dispersion
relations progressively deviates from the analytic results with
increasing $\kx H$. When $\kx H \lesssim 1$, however, the two agree
within $\sim4\%$ for $0.8 < \gamma < 2$. The agreement becomes
better for disks with smaller $A$. Since GI occurs for waves with
$\kx H \lesssim 1$ when self-gravity is included, this suggests that
$c_{\rm eff}$ given in equation \eqref{eq:rigid_approx} is a good
representative of the averaged sound speed.

\subsection{Free Boundary}

In the case of the free BCs, $\Sigma_{1,s}/\Sigma_{1,i} =
\left.\rho_{0a}\xi_{1z}|_{z=a}\middle/ {\int_0^a
\rho_{1,i}\,dz}\right.$. Then, the effective sound speed in equation
\eqref{eq:disp_wav} becomes
\begin{equation}\label{eq:c_eff_free}
c_{\rm{eff}}^2 = \left(
\frac{\int_{-a}^{a}\rho_0h_1/c_s^2\,dz}{\int_{-a}^{a}\rho_0h_1\,dz}
+ \frac{\rho_{0a}h_{1a}}{g_{0a}\int_0^a\rho_0h_{1}\,dz}
\right)^{-1},
\end{equation}
where equation \eqref{eq:bc_free} is used.  Equation
\eqref{eq:c_eff_free} indicates that waves in a pressure-confined
disk with free boundaries make use of two restoring forces: thermal
pressure and vertical gravity. The former drives acoustic waves,
while the latter is responsible for surface-gravity waves.  We term
these mixed waves acoustic-surface-gravity waves. The fact that the
effective sound speed has two terms, each arising from sound waves
and surface-gravity waves, is completely analogous to longitudinal
waves propagating in a distensible tube where the effective
compressibility of a gas is the sum of the true gas compressibility
and the distensibility of the tube (e.g., \citealt{Lighthill78}).

%% Figure3: Acoustic-gravity waves under the free BC
\begin{figure}
\epsscale{1.2}\plotone{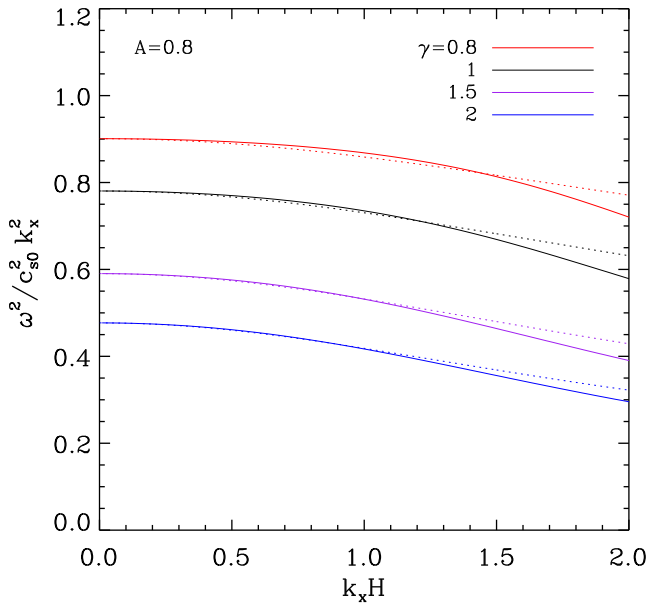} \caption{ Dispersion
  relations of the fundamental acoustic-surface-gravity modes in
  non-self-gravitating, non-rotating, polytropic disks with $\gamma =
  0.8, 1, 1.5, 2$. The case with mild pressure-confinement with
  $A=0.8$ ($\pext/\pi G\Sigma_0^2 = 0.28$) is shown. The frequency
  $\omega$ is normalized by $c_{s0}\kx$. Solid lines plot the full
  numerical results, while dotted lines draw the approximate
  dispersion relation \eqref{eq:c_eff_approx}.  }\label{fig:fund_free}
\end{figure}

Since ${\rho_{0a}} =\rho_{00} (1 - A^2)^{1/\gamma}$ and $g_{0a} =
4\pi G\rho_{00}AH_0$, it is apparent that the acoustic term in
equation \eqref{eq:c_eff_free} dominates for $A$ close to unity. On
the other hand, the surface-gravity term becomes important in disks
with $A \ll1 $. Appendix \ref{app:non-self} shows that in highly
confined disks, even fundamental modes have a particular solution
$h_1 \propto \cosh(\kx z)$. Equation \eqref{eq:c_eff_free} is then
reduced to
\begin{equation}\label{eq:c_eff_approx}
c_{\rm{eff}}^2 = \left({\frac{1}{\widetilde{c_s}^2} +
\frac{{\rho_{0a}}/{\rho_{00}}}{g_{0a}H}\frac{\kx H}{\tanh(\kx
H)}}\right)^{-1} =w\widetilde{c_s}^2, \,\,\,\mbox{for free BC},
\end{equation}
with $w$ given by equation (\ref{eq:w}). The sonic contribution
$\widetilde{c_s}$ is valid for $\kx H\lesssim1$, as in the rigid BC
case. The contribution from surface-gravity waves accounts for the
reduced density at the surfaces. In the limit of $A \ll1$, equations
\eqref{eq:disp_wav} and \eqref{eq:c_eff_approx} recover the usual
dispersion relation of surface-gravity waves
\begin{equation}\label{eq:disp_grav}
\omega^2 = g_{0a}\kx\tanh(\kx H),
\end{equation}
in an incompressible medium (e.g., \citealt{Goldreich+65a}).

Although equation \eqref{eq:c_eff_approx} is derived for a disk with
either $A \ll 1$ or $A\sim1$, we find that it is applicable for
arbitrary $A$ as long as $\kx H\lesssim1$. Figure
\ref{fig:fund_free} compares equation \eqref{eq:c_eff_approx}
(dotted lines) with the direct numerical solutions (solid lines)
found by solving the full perturbation equations (with self-gravity
and rotation neglected) for moderately confined disks with $A=0.8$
and $\gamma = 0.8,\,1,\,1.5,\,2$. The discrepancies between the two
are less than 0.5\% for $\kx H \lesssim 1$. This proves that
$c_{\rm{eff}}$ given in equation \eqref{eq:c_eff_approx} is
excellent in describing the phase velocity of the
acoustic-surface-gravity waves in the long-wavelength regime.

%% Figure4: w_0
\begin{figure}
\epsscale{1.2}\plotone{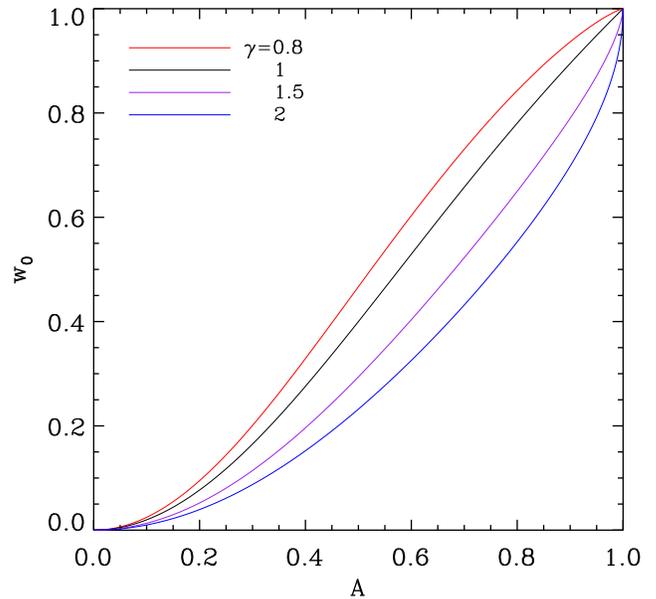} \caption{Ratio of the perturbed
  surface density inside the disk ($\Sigma_{1,i}$) to the total
  ($\Sigma_{1,i} + \Sigma_{1,s}$) for acoustic-surface-gravity waves
  in the long wavelength limit as a function of $A$ for differing
  $\gamma$.}\label{fig:w_0}
\end{figure}

From equation \eqref{eq:c_eff_approx}, it is now clear that the
variable $w$ defined by equation \eqref{eq:w} is a weight factor
representing the relative importance of acoustic to surface-gravity
terms in the dispersion relation of acoustic-surface-gravity waves.
In the long-wavelength limit, equation \eqref{eq:c_eff_free} leads
to
\begin{equation}\label{eq:w_0}
w_0 \equiv w|_{\kx \rightarrow 0} = \left[ 1 +
\gamma\alpha\frac{(1-A^2)^{1/\gamma}}{2A^2}\right]^{-1}.
\end{equation}
When $\gamma=1$, $w_0 = 2A^2/(1 + A^2) = (1 + \pext/\pi
G\Sigma_0^2)^{-1}$. Figure \ref{fig:w_0} plots $w_0$ as functions of
$A$ for disks with $\gamma = 0.8,\,1,\,1.5,\,2$.  As expected,
surface-gravity modes dominate for smaller $A$ and larger $\gamma$.
From equations \eqref{eq:w} and \eqref{eq:w_0}, one finds that the
mean amplitude of density perturbation inside the disk is given by
\begin{equation}\label{eq:amplitude}
\frac{\Sigma_{1,i}}{\Sigma_0} = \frac{2A^2}{\gamma\alpha}
\frac{\xi_{1z}\bigl|_{z=a}}{H},
\end{equation}
while $\Sigma_{1,s}/\Sigma_0 =
(1-A^2)^{1/\gamma}\xi_{1z}\bigl|_{z=a}/H$ at the disk surfaces.
Therefore, $|\Sigma_{1,i}/\Sigma_{1,s}|\ll1$ for small $A$. This
clearly demonstrates that the fundamental mode and its unstable
version (upon inclusion of self-gravity) are incompressible in
highly pressure-confined disks (e.g.,
\citealt{Elmegreen+78,Lubow+93,Nagai+98,Umekawa+99}).

\section{Gravitational Instability}\label{sec:gi}

In the preceding section, we derive expressions for the effective
sound speed for fundamental acoustic or acoustic-surface-gravity
waves in non-self-gravitating and non-rotating disks. In this
section, we derive the gravity reduction factors caused by finite
disk thickness that explain the numerical dispersion relations of GI
fairly well. We begin by considering non-rotating, zero-temperature
disks, and then include the effects of thermal pressure and
rotation.

\subsection{Pressureless Disks}\label{sec:JIDI}

Finite disk thickness is known to stabilize the conventional Jeans
instability by diluting self-gravity at the disk midplane.  As
mentioned in Introduction, $\mathcal{F}= (1+\kx H_0)^{-1}$ is often
used as the gravity reduction factor for an unbounded, exponential
disk with scale height $H_0$ (e.g., \citealt{Elmegreen82,
Elmegreen87, Kim+02}). In this subsection, we derive reduction
factors for the conventional Jeans modes and the distortional modes
in truncated polytropic disks.

We first consider a pressureless, non-rotating disk (i.e.,
$\widehat{h}_1 = \kappa_0^2 = 0$) for simplicity.  For given $\rho_1$,
equation \eqref{eq:per3} has a formal solution
\begin{equation}\label{eqa:psi1}
\psi_1(z) = - \frac{2\pi G}{\kx} \int
\rho_1(z^{\prime})e^{-\kx|z-z^{\prime}|}\,dz^{\prime},
\end{equation}
\citep[e.g.,][]{Kim+07}.  Substituting equation \eqref{eqa:psi1} into
equation \eqref{eq:omegasq_int}, one obtains
\begin{equation}\label{eq:growth}
\omega^2 = -2\pi G\Sigma_0 \kx \mathcal{F}(\kx),
\end{equation}
where
\begin{equation}\label{eq:reduction}
\mathcal{F}(\kx) = \frac{1}{\Sigma_0\Sigma_1}\iint \rho_0(z)\rho_1
(z^{\prime})e^{-\kx|z-z^{\prime}|}\,dzdz^{\prime},
\end{equation}
is the generalized reduction factor of self-gravity at wavenumber
$\kx$. It is apparent from equation \eqref{eq:reduction} that
$\mathcal{F} = 1 + O(\kx H)$ in the long wavelength limit,
regardless of density distributions.

Using equations \eqref{eq:sigma1s} and \eqref{eq:w}, we decompose
equation \eqref{eq:reduction} into two parts as
\begin{equation}\label{eq:reduction_decomp}
\mathcal{F}(\kx) = w \mathcal{F}_{\rm{J}}(\kx) + (1 -
w)\mathcal{F}_{\rm{D}}(\kx),
\end{equation}
where
\begin{equation}\label{eq:reduction_Jeans}
\mathcal{F}_{\rm{J}}(\kx) = \frac{1}{\Sigma_0\Sigma_{1,i}}\iint
\rho_0(z)\rho_{1,i}(z^{\prime})e^{-\kx|z-z^{\prime}|}\,dzdz^{\prime},
\end{equation}
and
\begin{equation}\label{eq:reduction_dist}
\mathcal{F}_{\rm{D}}(\kx) = \frac{e^{-\kx a}}{\Sigma_0}\int
\rho_0(z)\cosh(\kx z)\,dz,
\end{equation}
represent the reduction factors of self-gravity for
the conventional Jeans modes and distortional modes, respectively.

%% Figure5: Reduction factors
\begin{figure}
\epsscale{1.15} \plotone{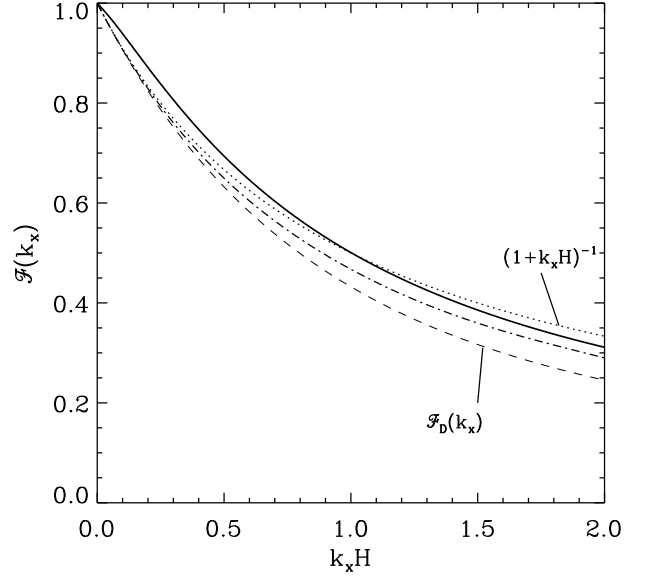} \caption{Various gravity reduction
  factors as functions of $\kx H$. The solid line plots
  $\mathcal{F}_{\rm{J}}$ by taking $\rho_0 \propto
  \mathrm{sech}^2(z/H)$ and $\rho_1$ from the eigensolutions of
  equations \eqref{eq:per_a}--\eqref{eq:per_d}, while the dot-dashed
  line gives $\mathcal{F}_{\rm{J}}$ by assuming
  $\rho_1\propto\rho_0\propto\mathrm{sech}^2(z/H)$.  The reduction
  factor $\mathcal{F}_{\rm{D}}$ (eq.\ [\ref{eq:FD}]) of the
  distortional modes is plotted as a dashed line. For comparison, the
  commonly used $\mathcal{F}=(1+\kx H)^{-1}$ is plotted as a dotted
  line.  }\label{fig:red}
\end{figure}

%% Figure6: Non-rotating dispersion relations comparison
\begin{figure*}
\plotone{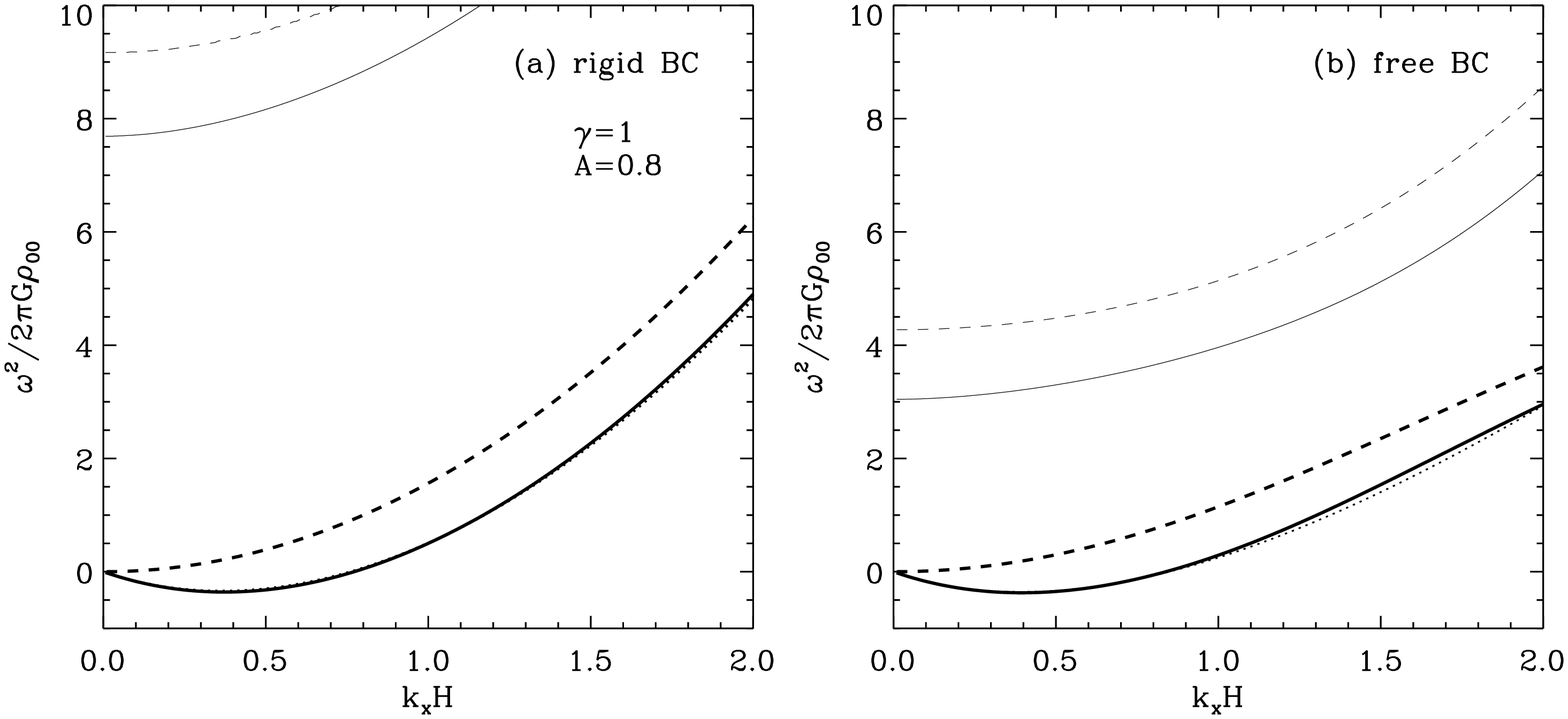} \caption{Dispersion relations of the
  fundamental modes (thick lines) and the first harmonics (thin lines)
  of the (a) acoustic and (b) acoustic-surface-gravity waves in disks
  with $A=0.8$ and $\gamma=1$.  Solid and dashed lines plot the
  numerical results with and without self-gravity, respectively, while
  dotted lines draw the approximate dispersion relation
  (eq.\ [\ref{eq:nonrot_anal}]) for the fundamental
  modes. }\label{fig:disp_comp}
\end{figure*}

We integrate equation \eqref{eq:reduction_Jeans} for an unbounded
($A=1$), isothermal disk with $\rho_0 \propto \sech^2(z/H_0)$. The
resulting $\mathcal{F}_{\rm{J}}$ with $\rho_{1,i} \propto \rho_0$ is
plotted in Figure \ref{fig:red} as a dot-dashed line, while the case
with $\rho_{1,i}$ numerically obtained from the full solution of
equations \eqref{eq:per_a}--\eqref{eq:per_d} is shown as a solid
line. Note that both are closely approximated by $(1+\kx H)^{-1}$
shown as a dotted line.  For pressure-confined disks, we empirically
found by varying $A$ as well as the BCs that
\begin{equation}\label{eq:FJr}
\mathcal{F}_{\rm{J}}=\frac{1}{1+\kx HA^{1/2}},  \;\;\text{for rigid
BC},
\end{equation}
and
\begin{equation}\label{eq:FJf}
\mathcal{F}_{\rm{J}}=\frac{1}{1+\kx H},  \;\;\text{for free BC},
\end{equation}
match the numerical dispersion relations of GI quite well (see
below).\footnote{For an unbounded, exponential disk with
$\rho_0\propto \rho_{1,i} \propto \exp(-|z|/H)$, equation
\eqref{eq:reduction_Jeans} yields
\begin{equation*}\label{eq:FJexp}
\mathcal{F}_{\rm{J}}^{\rm exp}=\frac{1}{1+\kx H} - \frac{\kx
H}{2(1+\kx H)^2},
\end{equation*}
which is different from the often-used expression (eq.\
[\ref{eq:FJf}]).  The latter is in fact the exact force correction
term \emph{at} the $z=0$ plane \citep[e.g.,][]{Kim+07}, while the
former is the correction factor \emph{averaged} along the
$z$-direction. Nevertheless, equation \eqref{eq:FJf} matches the
reduction factor for $\sech^2(z/H)$ disks better than
$\mathcal{F}_{\rm{J}}^{\rm exp}$.} The difference between equations
\eqref{eq:FJr} and \eqref{eq:FJf} is due to the fact that
$\rho_{1,i}$ in equation \eqref{eq:reduction_Jeans} is affected by
the BCs.

For small $A$, $\mathcal{F}$ is mostly from $\mathcal{F}_{\rm{D}}$
(see eq.\ [\ref{eq:reduction_decomp}]). In this case, the
distortional modes resulting from $\rho_{1,s}$ dominate, and it is
reasonable to take $\rho_0 \approx \mathrm{constant}$ and $a \approx
H$ in the evaluation of $\mathcal{F}_{\rm{D}}$. Equation
\eqref{eq:reduction_dist} then becomes
\begin{equation}\label{eq:FD}
\mathcal{F}_{\rm{D}}= e^{-\kx a}\frac{\sinh \kx H }{\kx H},
\end{equation}
which is plotted as a dashed line in Figure \ref{fig:red}.
Incidentally, $\mathcal{F}_{\rm{D}}$ is not much different from
$(1+\kx H)^{-1}$, either.

\subsection{Non-rotating Disks}\label{sec:wP}

We now explore GI of non-rotating, pressure-confined disks with the
effect of thermal pressure included. Combining the results of
Sections \ref{sec:fun} and \ref{sec:JIDI}, we write an approximate
dispersion relation
\begin{equation}\label{eq:nonrot_anal}
\omega^{2} = c_{\rm{eff}}^2\kx^2 - 2\pi G\Sigma_0k_x
(w\mathcal{F}_{\rm{J}} + (1-w)\mathcal{F}_{\rm{D}}),
\end{equation}
where $c_{\rm{eff}}$ is given by either equation
\eqref{eq:rigid_approx} or equation \eqref{eq:c_eff_approx}
depending on the BCs.

%% Figure7: GI Jeans mode
\begin{figure*}
\epsscale{1}\plotone{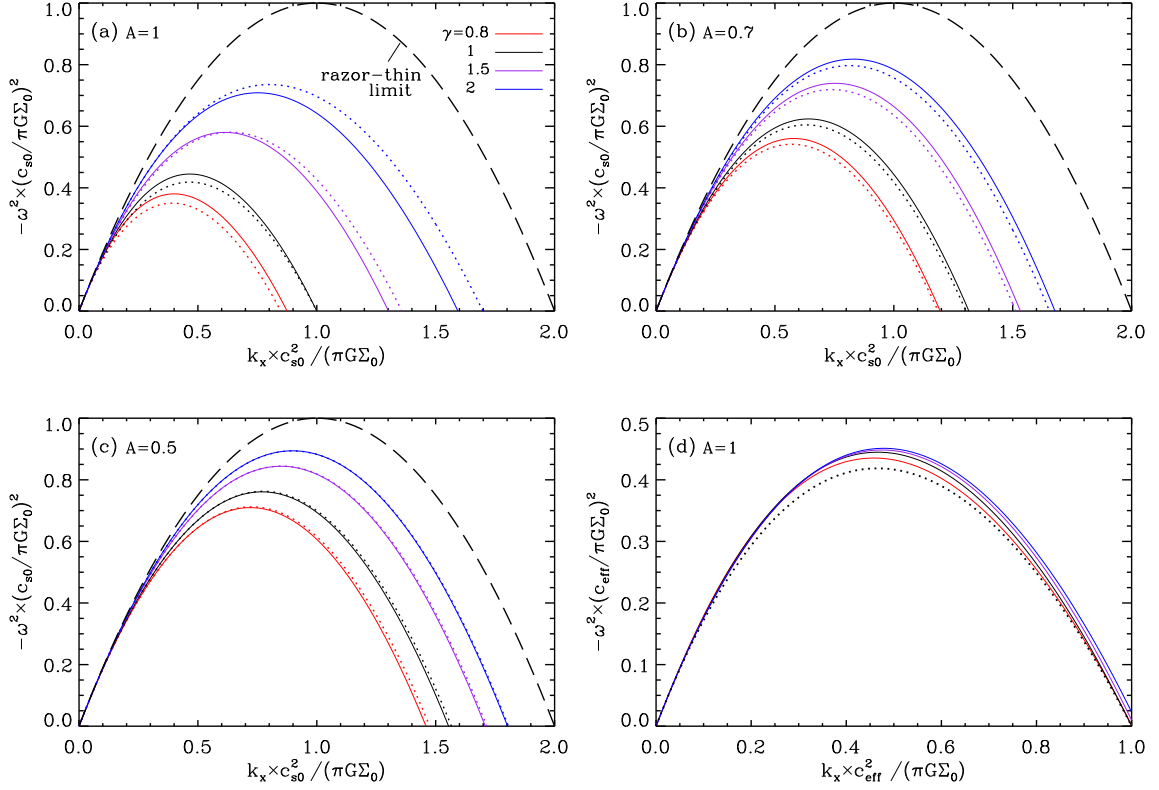}
\caption{ (a)--(c): Dispersion relations of the pure Jeans modes in
  non-rotating, pressure-confined, polytropic disks under the rigid
  BCs. The wavenumber and growth rate are normalized by using the
  midplane sound speed. The solid lines are the numerical results,
  while the approximate dispersion relations
  (eq. [\ref{eq:nonrot_anal}]) are plotted as dotted lines.  The
  razor-thin dispersion relation ($A \rightarrow 0$) is shown as a
  long dashed line in each panel.  (d): The $A=1$ case plotted in (a)
  is redrawn using $c_{\rm{eff}}$ in the normalization of $\omega$ and
  $k_x$. }\label{fig:GI_rigid}
\end{figure*}

\subsubsection{Rigid Boundary}

%% Figure8: GI mixed mode
\begin{figure*}
\epsscale{1}\plotone{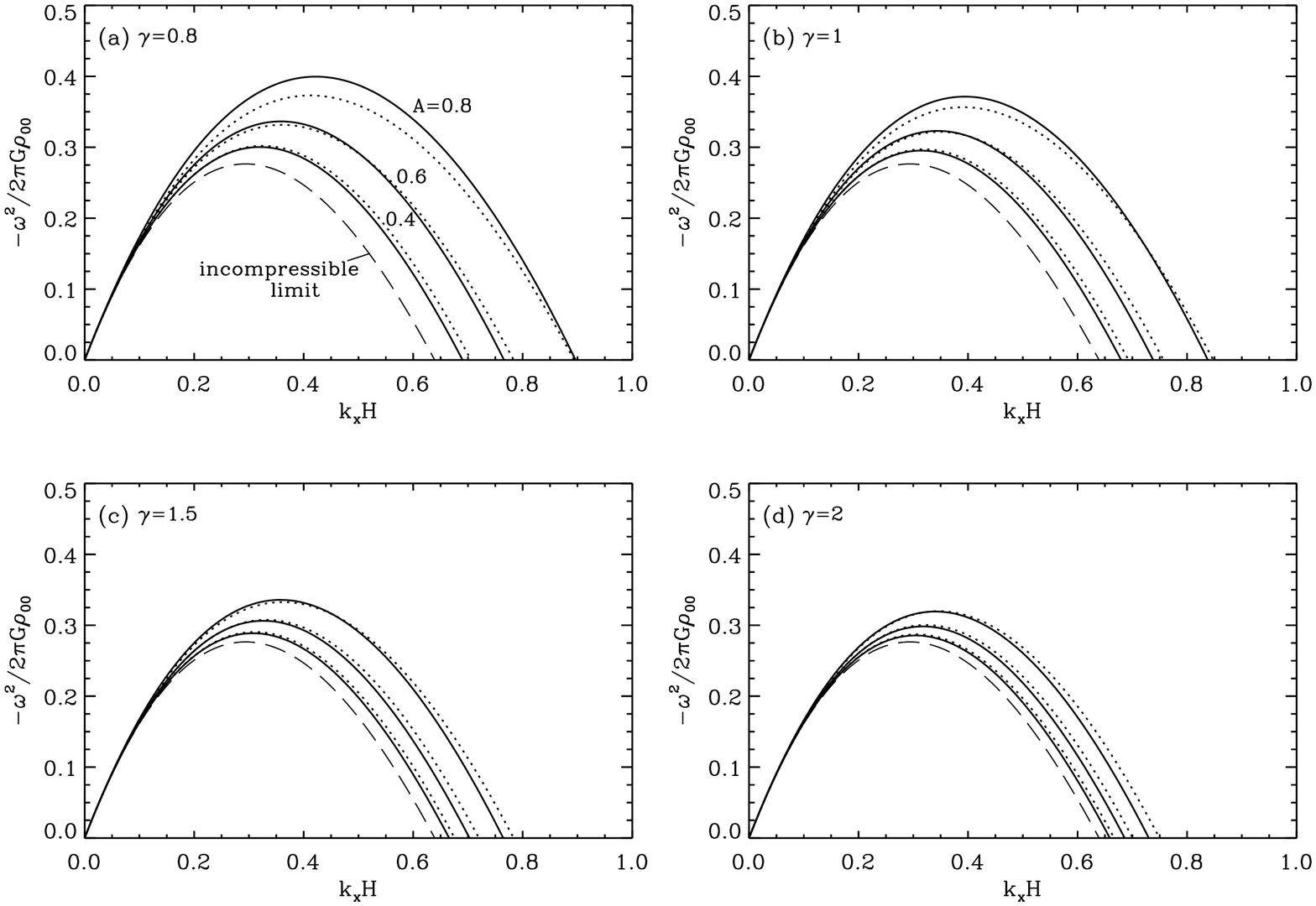}\caption{ Dispersion relations of the
  mixed GI in non-rotating, pressure-confined polytropic disks with
  $\gamma=0.8, 1, 1.5, 2$ and $A=0.4, 0.6, 0.8$ under the free BCs.
  Solid lines draw the numerical results, while dotted lines plot the
  approximate dispersion relation (eq.\ [\ref{eq:nonrot_anal}]).  In
  each panel, the pure distortional modes in the limit of $A
  \rightarrow 0$ are shown as a long dashed line.}\label{fig:GI_free}
\end{figure*}

First, we impose the rigid BCs with $w = 1$.  Figure
\ref{fig:disp_comp}a plots the dispersion relations of waves with
(solid lines) and without (dashed lines) self-gravity in disks with
$\gamma = 1$ and $A=0.8$ for the fundamental acoustic modes (thick)
as well as the first harmonics (thin) obtained numerically by
solving the perturbation equations
\eqref{eq:per_a}--\eqref{eq:per_d}. The approximate dispersion
relation \eqref{eq:nonrot_anal} for the fundamental mode is plotted
as dotted lines, which are in good agreement with the numerical
results for $\kx H\simlt 2$. Clearly, self-gravity lowers the
frequency and it is the fundamental mode that becomes unstable upon
inclusion of self-gravity. GI under the rigid BCs is the
conventional Jeans instability since surface distortion is absent.

Figure \ref{fig:GI_rigid} plots as solid lines the numerical
dispersion relations of pure Jeans modes for some selected values of
$A$ and $\gamma$, which agree quite well with equation
\eqref{eq:nonrot_anal} drawn as dotted lines: the relative
differences between the true and approximate values of the maximum
growth rates and critical wavenumbers are less than 5\% for $0.8 \le
\gamma \le 2$ and $0 \le A \le 1$. For fixed $\Sigma_0$ and
$c_{s0}$, the Jeans modes are more unstable for smaller $A$ because
the disk becomes geometrically thinner.  In the limit of $A
\rightarrow 0$, the dispersion relations of GI in pressure-confined
disks become identical to those of an infinitesimally-thin disk with
the same $\Sigma_0$ and $c_{s0}$, shown as dashed lines in Figure
\ref{fig:GI_rigid}a-c. For fixed $A$, disks with larger $\gamma$ are
more unstable owing to a steeper temperature gradient as well as a
smaller vertical scale height ($H_0 = Ac_{s0}^2/ (\pi
G\Sigma_0\gamma) \propto \gamma^{-1}$). Figure \ref{fig:GI_rigid}d
replots Figure \ref{fig:GI_rigid}a by scaling $\omega$ and $\kx$ in
terms of $c_{\rm{eff}}$ rather than the midplane sound speed
$c_{s0}$. Close agreement among the curves with different $\gamma$
evidences that the effective sound speed defined in equation
\eqref{eq:rigid_approx} is really a good representative of the mean
averaged sound speed of polytropic disks.

\subsubsection{Free Boundary}

Under the free BCs, perturbations experience two types of restoring
forces: compressibility and surface gravity. The former and latter
dominate when $A\sim1$ and $A\ll1$, respectively. For general $A$,
these are mixed together to become acoustic-surface-gravity waves.
Figure \ref{fig:disp_comp}b plots the numerical (solid lines) and
approximate (dotted lines; eq.\ [\ref{eq:nonrot_anal}]) dispersion
relations  for the fundamental modes (thick) and the first harmonics
(thin) in disks with $A=0.8$ and $\gamma=1$. Non-self-gravitating
waves are plotted as dashed lines, while solid lines draw the
self-gravitating waves. The close agreement between the numerical
and analytic results shows that the fundamental modes under the free
BCs are really acoustic-surface-gravity waves that become unstable
when self-gravity is included. In what follow, we call the unstable
acoustic-surface-gravity modes the mixed GI.
Figure \ref{fig:GI_free} plots as solid lines the numerical
dispersion relations of the acoustic-surface-gravity modes in
non-rotating, pressure-confined disks with various values of $A$ and
$\gamma$. The wavenumber and growth rate are normalized by the
effective half-thickness $H$ and the free-fall time at the midplane,
$(2\pi G\rho_{00})^{-1/2}$, respectively. The cases with $\gamma =
1$ recover the results of earlier studies (e.g., \citealt{Simon65,
Elmegreen+78,Usami+95,Iwasaki+11}). Shown also as dotted lines are
the approximate dispersion relation in good agreement with the
numerical results. Regardless of $A$, the maximum growth rate,
$|\omega_{\rm{max}}|$, of unstable modes occurs at
$kH\sim0.3$--$0.4$. However, the physical nature of GI is markedly
different depending on the degree of pressure confinement. When the
external pressure is weak ($A \approx 1$), unstable modes are
dominated by acoustic-type perturbations ($w\sim1$), making the
mixed GI under the free BCs similar to the pure Jeans mode occurring
under the rigid BCs. When perturbations from surface distortion
dominate ($A\ll1$), on the other hand, equation
\eqref{eq:nonrot_anal} is reduced to
\begin{equation}\label{eq:disp_dist}
\omega^2 = g_{0a}\kx\tanh(\kx H) -2\pi G\Sigma_0\kx \times \frac{1 -
e^{-2\kx H}}{2\kx H},
\end{equation}
for the pure distortional instability (e.g.,
\citealt{Goldreich+65a}).

\begin{figure*}
\epsscale{1}\plotone{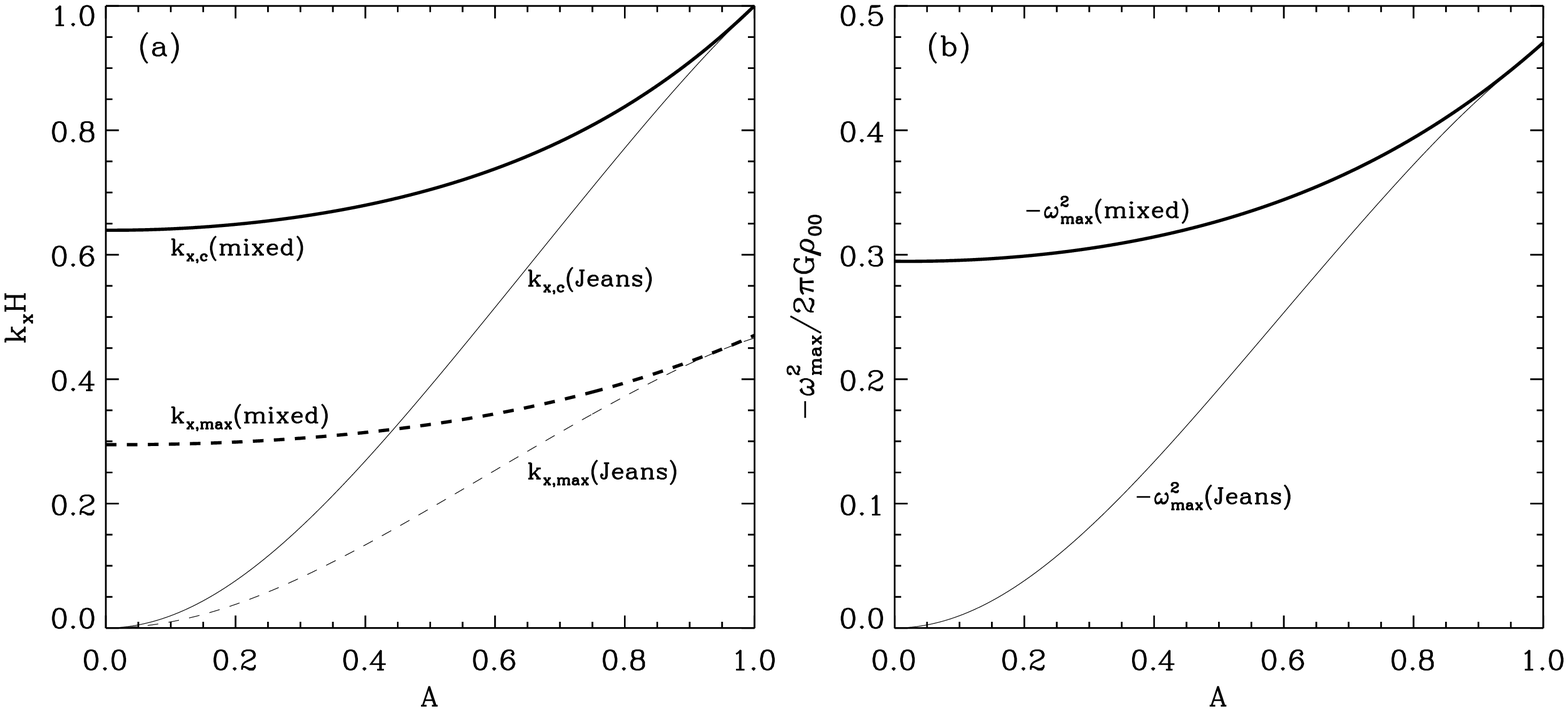} \caption{(a) Maximum ($k_{x,\rm max}$;
dashed lines) and critical
  ($k_{x,c}$; solid lines) wavenumbers, and (b) squared maximum growth
  rate as functions of $A$ for $\gamma=1$. The thick lines correspond
  to the mixed GI under the free BCs, while the thin lines are for
  the pure Jeans modes under the rigid BCs.  }\label{fig:GI_crit}
\end{figure*}

Figure \ref{fig:GI_crit} compares the marginal wavenumber $k_{x,c}$,
the most unstable wavenumber $k_{x,\rm max}$, and the maximum growth
rate $\omega_{\rm{max}}$ of the pure Jeans instability with those of
the mixed GI of isothermal disks as functions of $A$. The mixed GI
is identical to the pure Jeans mode when $A=1$, but the former in
general has a higher growth rate and occurs at a shorter wavelength
than the latter for arbitrary $A$. This is because the effective
sound speed varies with $A$ differently depending on the BCs. For
$A\ll1$, $c_{\rm eff} \sim c_{s0}$ for the pure Jeans modes (eq.\
[\ref{eq:alpha}]), while $c_{\rm eff} \sim \sqrt{g_{0a} a} \propto
A$ for the mixed GI (eq.\ [\ref{eq:c_eff_approx}]). Since both types
of GI have the characteristic wavenumber $k_{x,c}\sim \pi G
\Sigma_0/c_{\rm eff}^2$ and growth rate $|\omega| \sim
k_{x,c}/c_{\rm eff}$, the critical wavenumber and maximum growth
rate of the Jeans modes are smaller by a factor of $A^2$ and $A$,
respectively, compared to those of the mixed GI.

\subsection{Rotating Disks and Stability Criteria}\label{sec:Q}

\begin{figure}
\epsscale{1.1}\plotone{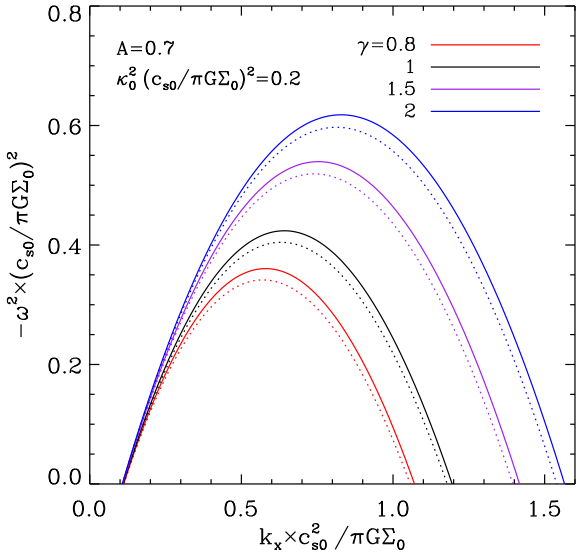} \caption{Dispersion relations of
rotating, polytropic disks with
  $A=0.7$ and $\gamma = 0.8,\,1,\,1.5,\,2$ when $\kappa_0^2 =
  0.2\,(\pi G\Sigma_0/c_{s0})^{2}$.
  Solid lines plot the numerical results, while dotted lines draw the
  approximate dispersion relation (eq.\ [\ref{eq:rot_approx}]).
}\label{fig:rot_disp}
\end{figure}

Finally, we include the effect of disk rotation and study the
stability condition to axisymmetric GI. In view of equation
\eqref{eq:omegasq_int}, one may assume that the gas motions arising
from thermal pressure and self-gravity are separable from the
epicycle motions, which is shown valid for $\kx H \simlt 1$ in the
preceding sections. The approximate dispersion relation of GI then
becomes
\begin{equation} \label{eq:rot_approx}
\omega^2 = c_{\rm{eff}}^2\kx^2 - 2\pi G\Sigma_0 \kx \mathcal{F}(\kx)
+ \kappa_0^2,
\end{equation}
which is a generalization of equation \eqref{eq:disp_razor} to
pressure-confined, polytropic disks with vertical stratification.
Figure \ref{fig:rot_disp} compares equation \eqref{eq:rot_approx}
shown as dotted lines with the full numerical results plotted as
solid lines for disks with $A=0.7$, $\gamma= 0.8,\,1,\,1.5,\,2$, and
$\kappa_0^2= 0.2(\pi G \Sigma_0/c_{s0})^2$, which are in good
agreement with each other, demonstrating again that equation
\eqref{eq:rot_approx} is accurate in describing GI of
pressure-confined polytropic disks provided $c_{\rm{eff}}$ and
$\mathcal{F}$ are chosen appropriately.

%% Figure9: Isothermal disk, max, critical

%% Figure10: Rotating disks' dispersion relations

We generalize Toomre's stability parameter originally defined for a
razor-thin disk to vertically-stratified, pressure-confined disks as
\begin{equation} \label{eq:toomre_def}
\Qeff \equiv \frac{c_{\rm{eff}}\kappa_0}{\pi G\Sigma_0}.
\end{equation}
Using $c_{\rm{eff}}/H_0 = (2\alpha\gamma\pi G\rho_{00})^{1/2}$ for
$A=1$, one can show that the stability parameter $Q_M=\Omega^2/(4\pi
G\rho_{00})$ used by \citet{Mamatsashvili+10} is equal to
$\sqrt{\Qeff/(8\alpha\gamma)}$. Since $\mathcal{F} \lesssim 1$, the
critical value of $Q_{\rm{eff}}$ is less than unity. Note that in
the limit of strong pressure confinement, $\Qeff$ is reduced to
\begin{equation}\label{eq:TooA0}
\Qeff \rightarrow \frac{\sqrt{g_{0a}a}\kappa_0}{\pi G\Sigma_0} =
\frac{\kappa_0}{\sqrt{\pi G\rho_{00}}}, \;\;\; \mbox{for } A
\rightarrow 0,
\end{equation}
applicable to the pure distortional modes.

%% Figure11: Toomre criteria
\begin{figure*}
\epsscale{0.9}\plotone{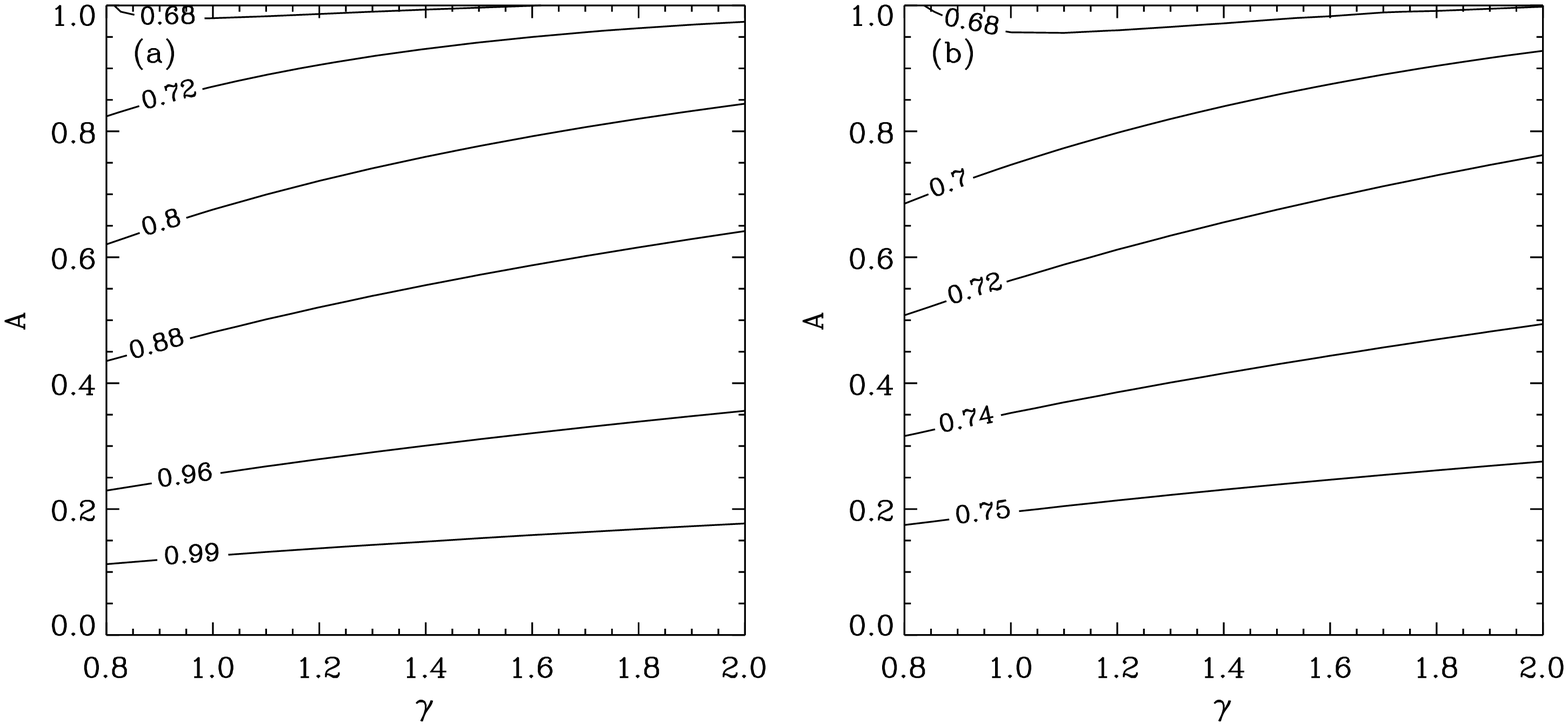}
\caption{Contours of the critical $Q_{\rm{eff}}$ values of (a) the
  pure Jeans modes under the rigid BCs and (b) the mixed GI under the
  free BCs for pressure-confined, polytropic disks. As $A\rightarrow 0$,
  $Q_{\rm{eff},c}$ approaches unity for the pure Jeans
  modes and $0.756$ for the mixed GI, independent of $\gamma$.}
\label{fig:toomre}
\end{figure*}

Figure \ref{fig:toomre} plots the critical values $\Qc$ for
axisymmetric GI in the ($\gamma$, $A$) plane. Figure
\ref{fig:toomre}a is for pure Jeans instability under the rigid BCs,
while the case of mixed GI with the free BCs is shown in Figure
\ref{fig:toomre}b. As $A \rightarrow 0$, $\Qc$ always converges to
$1$, for pure Jeans modes, corresponding to a razor-thin disk, and
to $0.756$ for the mixed GI, corresponding to an incompressible disk
\citep{Goldreich+65a}. In the pure Jeans instability, finite disk
thickness weakens self-gravity, decreasing $\Qc$ with increasing
$A$. For example, an unbounded isothermal disk has $\Qc = 0.676$
\citep{Goldreich+65a}. For the mixed GI, $\Qc \sim 0.68$--0.75
largely insensitive to $\gamma$.

\section{Summary and Discussion}\label{sec:discuss}

We investigate GI of rotating, pressure-confined,
vertically-stratified gas disks to axisymmetric perturbations. As an
initial equilibrium, we consider a self-gravitating, polytropic disk
in hydrostatic equilibrium. The disk is truncated by a constant
external pressure characterized by a dimensionless parameter $A$
(eq. [\ref{eq:A}]). To distinguish distortional modes of GI
occurring in highly confined disks from the conventional Jeans
modes, we adopt the rigid-surface BCs that allow only Jeans modes as
well as the free-surface BCs under which two modes coexist. Our
model disks do not include convective motions.

By deriving approximate dispersion relations and comparing them with
the numerical results from a set of the perturbation equations
(\ref{eq:per_a})--(\ref{eq:per_d}), we find that GI-unstable modes
are in general even-symmetry fundamental modes that have no vertical
node. Under the rigid BCs, the fundamental modes are simply acoustic
waves that propagate via thermal pressure only. Under the free BCs,
on the other hand, surface-gravity waves arising from surface
distortion provide an additional restoring force, forming
acoustic-surface-gravity waves. In the presence of self-gravity, the
acoustic waves become unstable to pure Jeans modes, while the
surface-gravity waves become unstable to incompressible,
distortional modes. Therefore, disks under the free BCs are unstable
to the mixed GI in which the Jeans and distortional modes coexist.
When pressure confinement is weak ($A\sim1$), the Jeans modes
dominate, while the distortional modes become prevalent in strongly
confined disks. The relative importance of the Jeans modes can be
measured by the dimensionless parameter $w_0$ defined by equation
\eqref{eq:w_0}.

While polytropic disks are vertically stratified both in density and
temperature, we find that the effective sound speed $c_{\rm{eff}}$
defined either by equation \eqref{eq:rigid_approx} for the rigid BCs
or equation \eqref{eq:c_eff_free} for the free BCs represents the
mean propagation speed of acoustic-surface-gravity waves quite well.
Under the rigid BCs, $c_{\rm{eff}}$ is given by the
density-weighted, harmonic mean of the local sound speeds, while it
has a contribution from the surface-gravity waves under the free
BCs.

We derive the reduction factors $\mathcal{F}(\kx)$ responsible for
the reduced gravity due to finite disk thickness for the pure Jeans
modes (eqs.\ [\ref{eq:FJr}] and [\ref{eq:FJf}]) and for the
distortional modes (eq.\ [\ref{eq:FD}]).  The approximate dispersion
relation \eqref{eq:rot_approx} matches the numerical results very
closely as long as the effective sound speed and the gravity
reduction factors are properly chosen. Disk rotation introduces
inertial waves associated with epicyclic motion, but it is again the
fundamental mode of acoustic-surface-gravity waves that are subject
to GI. Using the effective sound speed, we define the stability
parameter $\Qeff$ as in equation \eqref{eq:toomre_def} that
generalizes the Toomre's parameter into pressure-confined,
polytropic disks. The pure Jeans modes have the critical values
$\Qc=0.67$ -- 1 with the larger value corresponding to a razor-thin
disk, while the mixed GI has $\Qc = 0.67$ -- 0.76 insensitive to
$\gamma$.

As mentioned in Introduction, the origin of the mixed GI of
pressure-confined disks has been uncertain.  Various mechanisms such
as neutral modes \citep{Lubow+93}, pinching force
\citep{Elmegreen89, Umekawa+99, Wunsch+10}, and surface-gravity
waves \citep{Iwasaki+11} have been proposed as the cause of GI.  In
this paper, we clearly show that the mixed GI occurring under the
free BCs is a combination of pure Jeans and distortional modes, the
latter of which is generated by deformation of free surfaces. The
existence of an additional restoring force due to the vertical
gravity excites low-frequency fundamental modes, as opposed to
high-frequency acoustic modes. With a reduced effective sound speed,
the mixed GI has a larger growth rate and a smaller length scale
than the pure Jeans modes.

Recently, \citet{Iwasaki+11} performed a linear stability analysis
of expanding isothermal shells around \ion{H}{2} regions. They
considered combinations of the rigid and free BCs as well as an
asymmetric density profile caused by the bulk deceleration of a
shell. They found that the numerically calculated dispersion
relations are well described by an approximate relation $\omega^2 =
c_{{\rm eff},I}^2\kx^2- 2\pi G\Sigma_0\kx$, where $c_{{\rm
eff},I}^2=[B \times 4\pi G \Sigma_0 H +(c_s/2)^2]$. Although the
dimensionless parameter $B=0.39$ in their effective sound speed
$c_{{\rm eff},I}$ gives the best fit to their numerical growth
rates, its physical nature is somewhat ambiguous. Noting that
$H=A^2c_s^2/(\pi G\Sigma_0)$, the above relation can be arranged as
$\omega^2=c_{\rm eff}^2\kx^2 - 2\pi G\Sigma_0\kx \mathcal{F}_{\rm
I}(\kx)$, similarly to equation \eqref{eq:nonrot_anal}, with the
effective sound speed of acoustic-surface-gravity waves $c_{\rm
eff}=2A^2/(1+A^2)c_s^2$ in the long-wavelength limit (eq.\
[\ref{eq:c_eff_approx}]) and the gravity reduction factor modified
to $\mathcal{F}_{\rm I}(\kx)=1 - \kx H \times [2B +
(1-7A^2)/(8A^2(1+A^2))]$. For $A\gtrsim0.4$ (or, $\pext/(\pi
G\Sigma_0)\simlt 21/8$) considered by \citet{Iwasaki+11},
$\mathcal{F}_{\rm I}= 1 -(0.4$--$0.6)\kx H$ in reasonable agreement
with the reduction factors that we found (see Figure \ref{fig:red}).
This suggests that $c_{{\rm eff},I}$ in the approximate dispersion
relation of \citet{Iwasaki+11} contains a contribution from the
reduced gravity due to finite shell thickness.

While the mixed GI of pressure-confined disks has received a little
attention compared to pure Jeans modes, it may be responsible for
structure formation via fragmentation of thin shells in the ISM. For
example, \citet{Boyd+05} claimed that GI of shock-compressed layers
can explain the formation of Jupiter-mass (significantly smaller
than the Jeans mass) objects detected in young star clusters
\citep{Zapatero+02}. Recently, \citet{Wunsch+12} reported that the
observed slopes of the mass spectrum of molecular clouds in the
Carina Flare are indeed consistent with those expected from GI of
pressure-confined shells. Distortional modes in flattened systems
can be thought of as arising from the tendency to lower the
gravitational potential energy by transforming into a spherical
configuration \citep{Usami+95}. Thus, the mixed GI would occur not
only in a planar geometry but also in a cylindrical gaseous column.
In the presence of magnetic fields, distortional instability is
suppressed in the direction parallel to the magnetic fields
\citep{Nagai+98}, naturally leading to cloud formation in
filamentary shapes.

\acknowledgments We are grateful to the referee for a helpful
report. This work was supported by the National Research Foundation
of Korea (NRF) grant funded by the Korean government (MEST), No.\
2010-0000712.

\appendix

\section{Hydrostatic Equilibrium}\label{app:ini}

In this appendix, we derive an analytic solution of equations
(\ref{eq:HSE1}) and (\ref{eq:HSE2}) for the density distributions of
self-gravitating polytropic disks.  We begin by defining the
dimensionless variables
\begin{equation}\label{eq:xi}
\xi = z/H_0,
\end{equation}
\begin{equation}\label{eq:rho0}
\Theta(\xi) = (\rho_0/\rho_{00})^{1/n},
\end{equation}
where $n\equiv 1/(\gamma-1)$ is the polytropic index. Then,
equations (\ref{eq:HSE1}) and (\ref{eq:HSE2}) result in
\begin{equation}\label{eq:LE}
\frac{d^2\Theta}{d\xi^2} = - \left(\frac{2}{1+n}\right) \Theta^n.
\end{equation}
Equation (\ref{eq:LE}) is the usual Lane-Emden equation in the
planar geometry (e.g., \citealt{Viala+74a}), which can be solved
with the proper boundary conditions $\Theta(0)=1$ and
$d\Theta(0)/d\xi=0$.

Now, we combine equations (\ref{eq:HSE2}) and (\ref{eq:mu}) together
with equations (\ref{eq:xi}) and (\ref{eq:rho0}) to obtain
\begin{equation}\label{eq:mu2}
\frac{d}{d\xi} = {\Theta^n}\frac{d}{d\mu}.
\end{equation}
On the other hand, by integrating equation (\ref{eq:HSE2}) over $z$ and
using equation (\ref{eq:LE}), we express $\mu$ in equation (\ref{eq:mu}) as
\begin{equation}\label{eq:mu3}
\mu = -\left(\frac{1+n}{2}\right) \frac{d\Theta}{d\xi}
=- \left(\frac{n+1}{2}\right) \Theta^n \frac{d\Theta}{d\mu},
\end{equation}
where the second equality utilizes equation (\ref{eq:mu2}).
The solution of equation (\ref{eq:mu3}) that satisfies the boundary
conditions is simply
\begin{equation}\label{eq:dsol}
\Theta = ( 1 - \mu^2) ^{1/(n+1)},
\end{equation}
or
\begin{equation}\label{eq:dsol1}
\rho_0 (\mu)= \rhoo ( 1 - \mu^2) ^{1/\gamma},
\end{equation}
(see also, \citealt{Harrison+72}).

To find $\rho_0$ in terms of $z$, we substitute equations
(\ref{eq:xi}) and (\ref{eq:dsol}) into equation (\ref{eq:mu2})
to obtain
\begin{equation}\label{eq:zH0}
\frac{z}{H_0} = \int_0^\mu \frac{d\mu}{(1-\mu^2)^{1/\gamma}}.
\end{equation}
A formal solution of equation (\ref{eq:zH0}) is given by
\begin{equation}\label{eq:z_over_H}
\frac{z}{H_0}= {}_{2}F_{1}\left(
\case{1}{2},\,\case{1}{\gamma}\,;\case{3}{2}\,;\mu^2 \right)\mu,
\end{equation}
where ${}_{2}F_{1}$ denotes the Gaussian hypergeometric function.
Equation (\ref{eq:zH0}) is integrable for a few particular values of
$\gamma$:
\begin{enumerate}
% gamma=2/3
\item When $\gamma=2/3$,
\begin{equation}
{z}/{H_0} = {\mu}/\sqrt{1-\mu^2}, \;\;\;\;
{\rm and}\;\;\;
\rho_0= \rhoo \left( 1 + \left({z}/{H_0}\right)^2\right)^{-3/2},
\end{equation}
where $H_0^2=K/(2\pi G\rho_{00}^{4/3})$ (e.g., \citealt{Larson85}).
% gamma=1
\item When $\gamma=1$,
\begin{equation}\label{eq:iso_mu}
z/H_0 = \tanh^{-1} \mu, \;\;\;\;
{\rm and}\;\;\;
\rho_0= \rhoo \sech ^2(z/H_0),
\end{equation}
where $H_0^2=K/(2\pi G\rhoo)$ (e.g.,
\citealt{Ledoux51,Goldreich+65a}).
% gamma=2
\item When $\gamma=2$,
\begin{equation}
z/H_0 = \sin^{-1} \mu, \;\;\;\;
{\rm and}\;\;\;
\rho_0= \rhoo \cos(z/H_0),
\end{equation}
where $H_0^2=K/(2\pi G)$ (e.g., \citealt{Goldreich+65a}).
% gamma=infty
\item When $\gamma=\infty$,
\begin{equation}
z/H_0 = \mu, \;\;\;\;
{\rm and}\;\;\;
\rho_0= \rhoo,
\end{equation}
(e.g., \citealt{Harrison+72}).
\end{enumerate}

\section{Boundary Conditions}\label{app:bc}

Equations \eqref{eq:per_a}--\eqref{eq:per_d} possess a reflection
symmetry with respect to the $z=0$ plane, that is, invariant under
the transformations $z \rightarrow -z$, $\xi_{1z} \rightarrow
-\xi_{1z}$, $\psi_1^{\prime} \rightarrow -\psi_1^{\prime}$, $h_1
\rightarrow h_1$, $\psi_1 \rightarrow \psi_1$. Thus, perturbations
are in general a superposition of even-symmetry modes and
odd-symmetry modes. Of these, odd-symmetric modes satisfying
$\xi_{1z}(z) = \xi_{1z}(-z)$, $h_1(z) = -h_1(-z)$, $\psi_1(z) =
-\psi_1(-z)$, $\psi_1^{\prime}(z) = \psi_1^{\prime}(-z)$ are shown
to be stable against GI \citep[e.g.][]{Simon65,
Elmegreen+78,Mamatsashvili+10}.  Therefore, we in this work consider
only even-symmetry modes for which it is suffice to write only two
conditions
\begin{equation} \label{eq:symm}
\xi_{1z}(0) = \psi_1^{\prime}(0) = 0.
\end{equation}
The other two conditions on $h_1(0)$ and $\psi_1(0)$ are
automatically satisfied from equations
\eqref{eq:per_a}--\eqref{eq:per_d}.

At the boundaries truncated by the external pressure, we consider
two different types of BCs: the rigid and free conditions.  In the
rigid BC, the boundaries are assumed to be fixed with a vanishing
Lagrangian displacement
\begin{equation} \label{eq:bc_rigid}
\xi_{1z}\bigl\vert_{z=a} = 0, \;\;\text{for rigid BC},
\end{equation}
(e.g., \citealt{Voit88}). This condition may be appropriate to
describe boundaries of a thin layer, as produced by colliding
clouds, since the shocked interfaces are stable against distortion
\citep{Usami+95, Iwasaki+11}.

The free BC allows for a deformed Lagrangian surface on which
the pressure is maintained constant to its initial value $\pext$
\citep{Goldreich+65a, Elmegreen+78, Lubow+93, Iwasaki+11}.
Since $\pext = p(a + \xi_{1z}\bigl|_{z=a}) = p_{0}\bigl|_{z=a} + p_{1}\bigl|_{z=a}
+ (\xi_{1z}\, dp_0/dz)_{z=a}$ to first order,
one obtains
\begin{equation} \label{eq:bc_free}
h_{1}\bigl|_{z=a} = g_{0a}\xi_{1z}\bigl|_{z=a}, \;\;\text{for free BC},
\end{equation}
where $g_{0a}\equiv d\psi_0/dz\bigl|_{z=a} $ is the vertical gravity
at the disk boundary (e.g., \citealt{Goldreich+65a}).

In self-gravitating disks, another condition comes from applying the
divergence theorem to the perturbed Poisson equation at the interface.
Integrating equation \eqref{eq:per3} over the volume of a thin shell
placed at the interface gives
\begin{equation} \label{free_bc2}
-\kx \psi_1|_{z=a} - \frac{d\psi_1}{dz}\Bigl\vert_{z=a} = 4\pi
G\rho_0\xi_{1z}\bigl|_{z=a},
\end{equation}
where the relation $d\psi_1/dz|_{z=a+} = -\kx \psi_1|_{z=a}$ is used
since the gravitational potential outside the disk satisfies
Laplace's equation.

\section{Method of Numerical Integration}\label{app:num}

For a given set of disk parameters ($\gamma$, $A$, $\kappa_0^2$), we
wish to find a dispersion relation $\omega = \omega(\kx)$ that
satisfies the perturbations equations
\eqref{eq:per_a}--\eqref{eq:per_d} subject to the two BCs and two
even-symmetry constraints. Since the perturbation equations are
linear in terms of the independent variables $(\xi_{1z}, h_1,
\psi_1, \psi_1^\prime)$, we are free to choose one of the variables
arbitrarily. We thus fix $h_1|_{z=a} = 1$ (or $\xi_{1z}|_{z=a} =1$)
and take trial values for $\psi_1|_{z=a}$ and $\omega^2$, while
obtaining $\xi_{1z}|_{z=a}$ (or $h_1|_{z=a}$ ) and
$\psi_1^\prime|_{z=a}$ from equations
\eqref{eq:bc_rigid}--\eqref{free_bc2} depending on the adopted BCs.
We then integrate equations \eqref{eq:per_a}--\eqref{eq:per_d} from
$z=a$ to $z=0$ using the fourth-order Runge-Kutta-Gill method
\citep{Abramowitz+72}. At the $z=0$ plane, we check the
even-symmetry conditions \eqref{eq:symm}: if these are not
fulfilled, we go back to the disk boundary, change $\psi_1|_{z=a}$
and $\omega^2$ slightly based on the Newton-Raphson technique, and
repeat the integrations. Solutions with accuracy of $10^{-4}$, that
is, with $|\xi_{1z}(0)/\xi_{1z}(a/2)|,
|\psi^{\prime}_1(0)/\psi^{\prime}_1(a)| < 10^{-4}$ are obtained
typically within less than three iterations.

\section{Mode Classification and Effects of Self-gravity}\label{app:WKB}

Here, we first derive local dispersion relations of waves in the WKB
limit, and compare them with the numerical results. This will help
us verify that acoustic modes become unstable in the presence of
self-gravity. We then discuss our results in comparison with the
results of \citet{Mamatsashvili+10} who mistakenly claimed that
gravity makes inertial modes unstable.

\subsection{Local Analysis}

We seek for the solutions of equations
(\ref{eq:per_a})--(\ref{eq:per_d}) for local waves with $k_x, k_z
\gg H_0^{-1}$. Defining $k_z^2 \equiv -\chi_1^{-1}d^2\chi_1 /dz^2$
with $\chi_1=h_1+\psi_1$, it is straightforward to derive
\begin{equation}\label{eq:WKB}
\omega^4 - \omega^2(\beta c_s^2k^2 + \kappa_0^2) + \kappa_0^2\,\beta
c_s^2k_z^2 = 0,
\end{equation}
where $k^2 = k_x^2 + k_z^2$ and $\beta \equiv 1 + {\psi}_1/{h}_1 = 1
- 4\pi G\rho_0/(c_s^2 k^2)$ is the correction factor for
self-gravity (e.g., \citealt{Chandrasekhar61}). The corresponding
eigenfunctions are
\begin{equation}\label{eq:eigen}
\left(
\begin{array}{c}
\rho_1/\rho_0 \\
u_{1x}/c_s \\
u_{1z}/c_s
\end{array}\right)
= \mathcal{A} \left(
\begin{array}{c}
1 \\
- \beta\omega c_s k_x/(\omega^2-\kappa_0^2)  \\
- \beta c_s k_z/\omega
\end{array}\right)
e^{i(\omega t + k_x x + k_zz)},
\end{equation}
where $\mathcal{A}$ is an arbitrary constant.

For waves with $|\beta| c_s^2 k^2 \gg \kappa_0^2$ or $k \gg k_z$,
equation (\ref{eq:WKB}) yields two approximate solutions
\begin{align}
\omega_p^2 & = \beta c_s^2k^2 + \kappa_0^2, \label{eq:WKB1} \\
\omega_r^2 & = \kappa_0^2 \frac{\beta c_s^2k_z^2}{\beta c_s^2k^2 +
\kappa_0^2}. \label{eq:WKB2}
\end{align}
It can be shown that the group and phase velocities of the waves
associated with $\omega_p^2$ are parallel to each other, while those
with $\omega_r^2$ are perpendicular.  This demonstrates that the
former is acoustic waves ($p$ modes) boosted by epicycle motions,
while the latter is inertial waves ($r$ modes). Note that inertial
waves require non-vanishing $k_z$ for propagation.

%% Figure12: Rotating disk dispersion relation comparison
\begin{figure}
\plotone{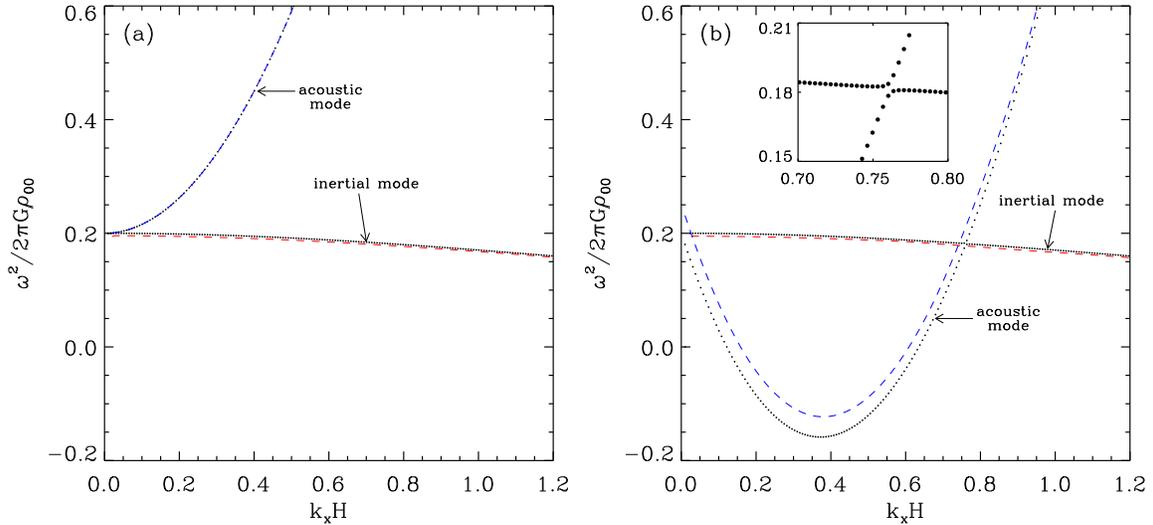} \caption{Comparison of the local
dispersion relations (dashed lines) with the numerical dispersion
relations (dots) for acoustic and inertial modes in (a)
non-self-gravitating and (b) self-gravitating cases.  For the
numerical results, an isothermal disk with $A=0.8$ and
$\kappa_0^2/(2\pi G\rho_{00})=0.2$ is taken and the rigid BCs are
adopted. In (a), the local dispersion relations are almost identical
to the numerical results. In (b), the acoustic and inertial modes
cross each other near $k_xH=0.77$, and the inset zooms in on the
crossing regions. While self-gravity changes the acoustic mode
dramatically, the inertial modes are almost intact. \label{fig:WKB}}
\end{figure}

It is apparent from equation \eqref{eq:WKB1} that the local acoustic
waves become gravitationally unstable provided self-gravity is
sufficiently strong with $\beta < -\kappa_0^2/(c_s^2k^2)$. On the
other hand, the inertial waves usually have $k_zH_0\gg1$ so that
$\omega_r^2 \approx \kappa_0^2 k_z^2/k^2$ regardless of $\beta$,
indicating that the character and frequencies of inertial waves are
almost unchanged by self-gravity. Figure \ref{fig:WKB} demonstrates
this by directly comparing equations (\ref{eq:WKB1}) and
(\ref{eq:WKB2}) shown as dashed lines with the numerical dispersion
relations plotted as dots obtained by integrating equations
(\ref{eq:per_a})--(\ref{eq:per_d}) subject to the rigid BCs.  An
isothermal disk with $A=0.8$ and $\kappa_0^2/(2\pi G\rho_{00}) =
0.2$ is considered for the numerical calculations using the method
described in Appendix \ref{app:num}. Only the results for the
fundamental acoustic modes (i.e., no node in the $z$-direction) and
the lowest-order inertial modes with a single node are shown. The
inset in Figure \ref{fig:WKB}b enlarges the regions with $0.7\leq
k_xH\leq 0.8$ and $0.15\leq \omega^2/(2\pi G\rho_{00})\leq 0.21$,
where the acoustic and inertial modes cross each other. In plotting
equations (\ref{eq:WKB1}) and (\ref{eq:WKB2}), we utilize the
eigenfunctions and their derivatives at $z=0$ to calculate $\beta$
and $k_z$. Note that agreement between the local and numerical
dispersion relations is excellent for the non-self-gravitating case.
Self-gravity makes $\beta\simlt -0.4$ for the acoustic modes, making
them unstable for $0.12<k_xH<0.63$. On the other hand, the inertial
modes have $\beta c_s^2 k_z^2/\kappa_0^2\simgt 38$ and thus remain
stable even in the presence of self-gravity.

As in our present work, \citet{Mamatsashvili+10} also calculated
dispersion relations of waves in vertically-stratified, polytropic,
rotating disks. Except for the internal gravity waves arising from
convective motions that are missing in our models,  their numerical
results are qualitatively similar to ours.\footnote{Note that
\citet{Mamatsashvili+10} considered disks with $\gamma>1$ and
$\pext=0$, automatically truncated at $z=z_{\rm cut}$ (see eq.\
[\ref{eq:trun}]). With the imposed free BCs, they identified
surface-gravity waves ($f$ mode), although the destabilizing effect
of surface distortion is absent in their models.} In interpreting
their results, however, they argued that the influence of
self-gravity is strongest on the inertial mode and that GI results
from the first harmonics of the inertial mode. This erroneous
finding seemingly results from the observation that the
long-wavelength unstable branch of the dispersion curves are
smoothly connected to the short-wavelength part of the inertial
modes, which occurs at $k_xH \simeq 0.77$ in Figure \ref{fig:WKB}b.
They claimed that the frequency ordering $\omega_r^2 \leq \kappa_0^2
< \omega_p^2 $ which is valid in the non-self-gravitating case holds
true also for the the self-gravitating counterpart, so that the
lowest-frequency modes are always the inertial modes regardless of
$k_x$. However, our local dispersion relations (eqs.\
\eqref{eq:WKB1} and \eqref{eq:WKB2}) show that this is correct only
when $\beta>0$, i.e., only for waves with relatively short
wavelength.\footnote{This is usually the case in stellar interiors
since $c_s^2\sim \pi G\rho_0 R^2$ from the condition of hydrostatic
equilibrium, with $R$ denoting the stellar radius. Then,
$\beta\simeq 1 - 0.1 (\lambda/R)^2>0$ for waves with wavelength
$\lambda<R$.} For gravitationally unstable modes with $\beta <
-\kappa_0^2/(c_s^2k^2)$, $\omega_p^2 < 0 <\omega_r^2 \leq
\kappa_0^2$. In Figure \ref{fig:WKB}b, the lower-frequency mode at
$k_xH <0.77$ is connected to the higher-frequency acoustic mode at
$k_xH > 0.77$, while the higher-frequency mode at $k_xH <0.77$ is
connected to the lower-frequency inertial mode at $k_xH > 0.77$.
When the inertial and acoustic modes are mixed together by having
similar frequencies and wavenumbers, it is quite ambiguous to tell
which one is which near $k_xH=0.77$.

\subsection{Comparison of Eigenfunctions}

A clear way to distinguish the modes is to compare the corresponding
eigenfunctions since they contain information on mode
characteristics. Inertial modes, being incompressible in nature,
should involve very weak density perturbations. This can be readily
seen from equation \eqref{eq:eigen} since
\begin{equation}
\frac{|\rho_1/\rho_0|}{|u_{1x}/c_s|} =
\frac{|\omega^2-\kappa_0^2|}{|\beta\omega| c_s k_x} \approx
\frac{\kappa_0 k_x}{\beta c_s k_z^2} \ll 1,
\end{equation}
for inertial modes with $k_x/k_z\ll 1$. On the other hand, acoustic
modes with $k_x/k_z\gg1$ have $|\rho_1/\rho_0|/|u_{1x}/c_s| \approx
c_s k_x /|\omega_p| \sim O(1)$, implying that they rely on
relatively large density perturbations.

%% Figure13: Rotating disk eigenfunctions comparison
\begin{figure}
\plotone{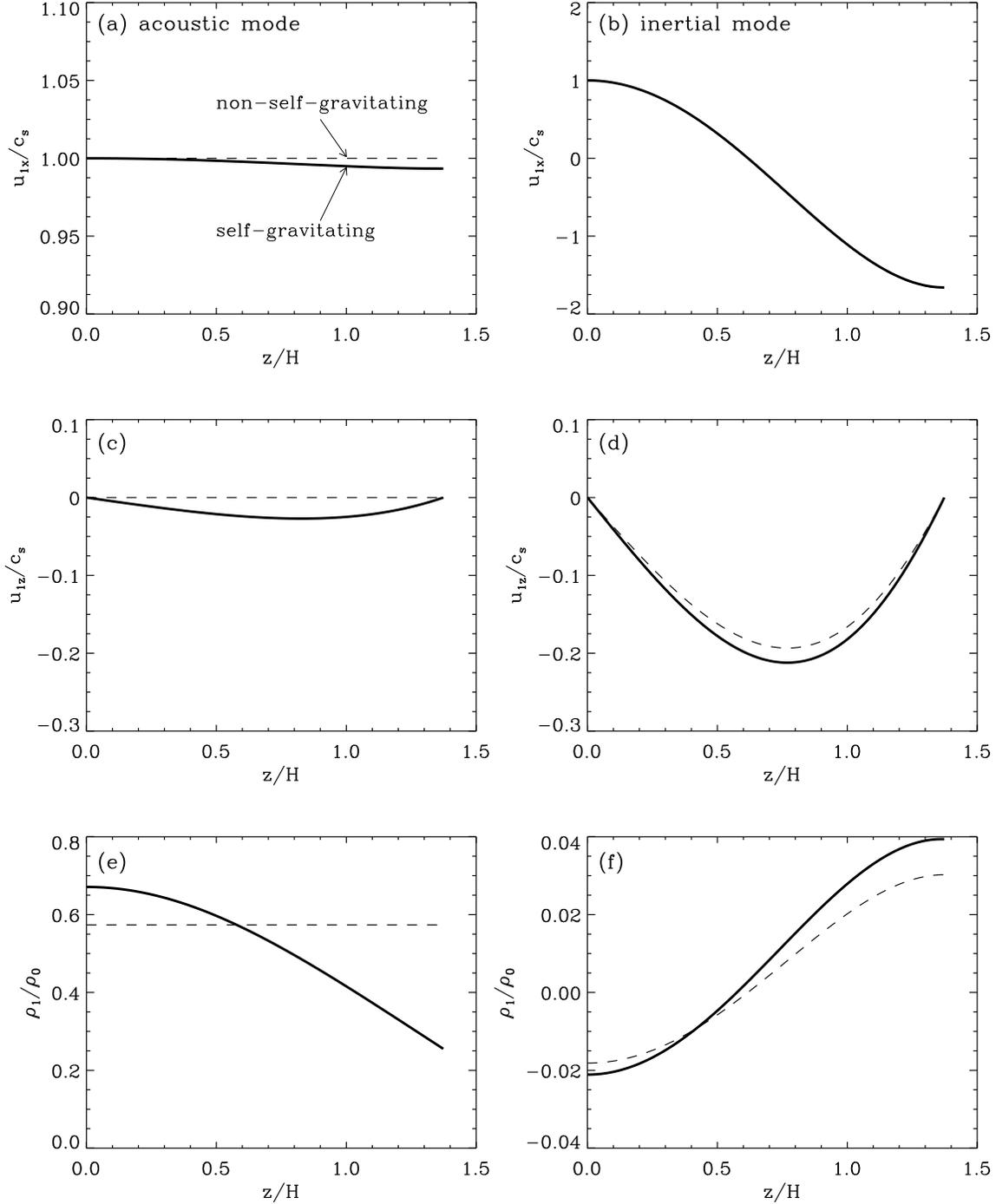} \caption{Comparisons of the
eigenfunctions for the acoustic modes (left panels) and the
lowest-order inertial modes (right panels) with $k_xH = 0.37$ shown
in Figure \ref{fig:WKB}. The solid and dashed lines correspond to
the self-gravitating and non-self-gravitating modes, respectively.
All eigenfunctions are normalized such that $u_{1x}=1$ at $z=0$.
Note that the inertial modes involve very little density
perturbations and are almost unaffected by self-gravity, while the
acoustic modes have $|u_{1z}/u_{1x}|\ll1$ since they propagate
primarily along the horizontal direction. \label{fig:eigen}}
\end{figure}

Figure \ref{fig:eigen} plots the vertical profiles of the
numerically-calculated eigenfunctions $u_{1x}$, $u_{1z}$, and
$\rho_1/\rho_0$ of the fundamental acoustic modes (left panels) and
the lowest-order inertial modes (right panels). The background state
is the same as in Figure \ref{fig:WKB}.  The most-unstable
wavenumber $k_x H=0.37$, well away from $k_xH=0.77$ where the two
modes are mixed, is chosen for all modes. The solid and dashed lines
correspond to the cases with and without self-gravity, respectively.
The normalization is such that $u_{1x}/c_s=1$ at $z=0$. The
non-self-gravitating acoustic mode has $\omega^2/(2\pi
G\rho_{00})=0.41$, $k_z=0$, $u_{1x}/c_s=1$, $u_{1z}=0$, and
$\rho_1/\rho_0=0.57$, independent of $z$, which are in fact exact
solutions (see Appendix \ref{app:non-self}). When self-gravity is
included, the acoustic mode becomes unstable with $\omega^2/(2\pi
G\rho_{00})=-0.16$; the corresponding eigenfunctions varies with $z$
only sightly with $k_zH=0.11$, but they do not have any node. For
acoustic modes, $|u_{1z}/u_{1x}| \ll1$ for $k_z/k_x\ll1$, consistent
with the prediction of the local analysis. The amplitude of the
perturbed density for the acoustic modes is of order unity relative
to the velocity perturbations. On the other hand, the inertial modes
have a node at around $z/H\sim 0.57$--$0.61$, regardless of the
presence of self-gravity. Gravity makes little change in $u_{1x}$
and small changes in $u_{1z}$ and $\rho_1$.  The corresponding
eigenfrequencies and vertical wavenumbers are $\omega^2/(2\pi
G\rho_{00})=0.1954, 0.1949$ and $k_zH=2.424, 2.421$ in
non-self-gravitating and self-gravitating cases, respectively. Note
that $|\rho_1/\rho_0|/|u_{1x}/c_s|\ll1$ for the inertial modes, just
as expected.

All of the above results suggest that the acoustic modes are
strongly affected by self-gravity to be unstable when self-gravity
is sufficiently strong, and that the inertial modes are not much
influenced by self-gravity and remain stable.

\section{Solutions for Waves in Non-self-gravitating Disks} \label{app:non-self}

When $\psi_1=\kappa_0=0$, equations \eqref{eq:per_a} and
\eqref{eq:per_b} become
\begin{align}
&\frac{d\xi_{1z}}{dz} = \frac{k^2}{\omega^2}h_1 -
\frac{1}{c_s^2}(h_1
  - g_0\xi_{1z}), \label{eq:nongrav_a} \\
&\frac{dh_1}{dz} = \omega^2\xi_{1z}. \label{eq:nongrav_b}
\end{align}
In equation \eqref{eq:nongrav_a}, the vertical gravity $g_0$ should
be considered as being external. Substituting equation
\eqref{eq:nongrav_b} into equation \eqref{eq:nongrav_a}, we obtain
\begin{equation}\label{eq:sturm}
\frac{d}{dz}\left(\rho_0\frac{d}{dz}h_1\right)
+\left(\frac{\omega^2}{c_s^2} -\kx^2\right)\rho_0h_1=0,
\end{equation}
where the equilibrium condition $d\rho_0/dz = -g_0\rho_0/c_s^2$ is
used. Note that equation \eqref{eq:sturm} together with the rigid
BCs and the symmetry constraint at the midplane is in the
Sturm-Liouville form with eigenvalue
\begin{equation}\label{eq:kzeigen}
k_z^2 \equiv {\omega^2}/{c_s^2} - k_x^2,
\end{equation}
indicating that the fundamental modes have the lowest frequency
(thus most susceptible to GI).

For a time being, we limit to an isothermal disk with $\rho_0
\propto \mathrm{sech}^2(z/H_0)$ for which analytic solutions of
equation \eqref{eq:sturm} can be derived. In terms of the
dimensionless variable $\mu$ (eq.\ [\ref{eq:iso_mu}]), equation
\eqref{eq:sturm} becomes
\begin{equation}
(1-\mu^2)\frac{d^2 v}{d\mu^2} -2\mu\frac{d v}{d\mu} + \left(2 -
\frac{\lambda^2}{1-\mu^2}\right)v =0\,,
\end{equation}
where $v\equiv (1-\mu^2)^{1/2}h_1$ and $\lambda^2 = 1 - (k_zH_0)^2$.
The above equation is of the associated Legendre type whose solution
is given by
\begin{equation}
v = C_1 P_1^{\lambda} + C_2 P_1^{-\lambda},
\end{equation}
where $C_1$ and $C_2$ are constants to be determined, and $P_1^{\pm
\lambda}$ is the Ferrers' associated Legendre function defined as
\begin{equation}
P_1^{\pm \lambda} = (\lambda \mp
\mu)\left(\dfrac{1+\mu}{1-\mu}\right)^{\pm \lambda/2},
\end{equation}
\citep{Whittaker+63}.

The even-symmetry condition, $dv/d\mu |_{\mu=0}$, at the midplane
requires $C_1 = C_2$. After some manipulations, we find $h_1$ in
terms of $z$ as
\begin{equation}\label{eq:h1}
h_1 =  \sinh(z/H_0)\sin(\widetilde{k_z}z)
-\widetilde{k_z}H_0\cosh(z/H_0)\cos(\widetilde{k_z}z),
\end{equation}
where $\widetilde{k_z} \equiv \sqrt{k_z^2 - H_0^{-2}}$ and the
proportionality constant $C_1$ or $C_2$ is omitted. The vertical
wavenumber that satisfies the rigid BC, $dh_1/dz|_{z=a}=0$, is then
given by
\begin{equation}\label{eq:kz2}
k_{z,n}^2 =\left\{\begin{array}{ll} 0, & \textrm{for fundamental
mode},
\\
{n^2\pi^2}/{a^2} + 1/{H_0^{2}}, & \textrm{for harmonics with order
$n=1, 2, 3, \cdots$.} \end{array}\right.
\end{equation}

Now we consider general polytropic disks, and seek for
fundamental-mode solutions of \eqref{eq:sturm}. While equation
\eqref{eq:sturm} cannot be solved in a closed form for arbitrary
$\gamma$, it has the simplest but important solution
\begin{equation}\label{eq:simple}
h_1 \rightarrow \mathrm{constant},
\end{equation}
in the long-wavelength limit regardless of the BCs. This corresponds
to equation \eqref{eq:h1} with $k_z=0$ for isothermal disks. For
waves with $\kx H \ll 1$, gas motions are restricted mostly to the
horizontal direction. In this case, the acceleration of the gas
along the vertical direction ($-\omega^2\xi_{1z}$ in eq.\
[\ref{eq:nongrav_b}]) becomes relatively unimportant, resulting in
$dh_1/dz \approx 0$. As Figure \ref{fig:acoustic} shows, $h_1
\approx \mathrm{constant}$ is reasonably good for $\kx H \simlt 1$.

Another limiting case is a strongly confined disk with $A \ll 1$,
for which $\rho_0(z) = \mathrm{constant}$ and $\omega^2/c_s^2 \ll
\kx^2$, the latter of which shall be verified a posteriori. Then,
the even-symmetry solution of equation \eqref{eq:sturm} is
\begin{equation}
h_1 \propto \cosh(\kx z),
\end{equation}
which corresponds to equation \eqref{eq:h1} in the limit of
$\widetilde{k_z}\rightarrow i\kx$ and $z/H_0\ll 1$. In this case,
$\nabla \cdot \boldsymbol{\xi_1} = 0$ and the system is essentially
incompressible. Note that equation \eqref{eq:disp_grav} gives
$\omega^2/c_s^2 < (g_{0a}H/c_s^2)\kx^2 \ll \kx^2$, as expected.

Up to now we have ignored the effect of disk rotation, but it is a
simple matter to show that the above results are valid also for
rotating disks provided $k_z$ in equation \eqref{eq:kzeigen} is
changed to
\begin{equation}
k_z^2 = \frac{\omega^2}{c_s^2} -
\frac{\omega^2}{\omega^2-\kappa_0^2}k_x^2\,,
\end{equation}
which is identical to equation \eqref{eq:WKB}. This implies that the
local solutions presented in Appendix \ref{app:WKB} are exact for
non-self-gravitating, isothermal disks.

%%\pagebreak


\begin{thebibliography}
\expandafter\ifx\csname natexlab\endcsname\relax\def\natexlab#1{#1}\fi

\bibitem[{{Abramowitz} \& {Stegun}(1972)}]{Abramowitz+72}
{Abramowitz}, M., \& {Stegun}, I.~A. 1972, {Handbook of Mathematical Functions}

\bibitem[{{Bell} {et~al.}(1997){Bell}, {Cassen}, {Klahr}, \&
  {Henning}}]{Bell+97}
{Bell}, K.~R., {Cassen}, P.~M., {Klahr}, H.~H., \& {Henning}, T. 1997, \apj,
  486, 372

\bibitem[{{Binney} \& {Tremaine}(2008)}]{Binney+08}
{Binney}, J., \& {Tremaine}, S. 2008, {Galactic Dynamics: Second Edition}
  (Princeton University Press)

\bibitem[{{Boley} {et~al.}(2007){Boley}, {Durisen}, {Nordlund}, \&
  {Lord}}]{Boley+07}
{Boley}, A.~C., {Durisen}, R.~H., {Nordlund}, {\AA}., \& {Lord}, J. 2007, \apj,
  665, 1254

\bibitem[{{Boley} {et~al.}(2006){Boley}, {Mej{\'{\i}}a}, {Durisen}, {Cai},
  {Pickett}, \& {D'Alessio}}]{Boley+06}
{Boley}, A.~C., {Mej{\'{\i}}a}, A.~C., {Durisen}, R.~H., {Cai}, K., {Pickett},
  M.~K., \& {D'Alessio}, P. 2006, \apj, 651, 517

\bibitem[{{Boss}(1997)}]{Boss97}
{Boss}, A.~P. 1997, Science, 276, 1836

\bibitem[{{Boyd} \& {Whitworth}(2005)}]{Boyd+05}
{Boyd}, D.~F.~A., \& {Whitworth}, A.~P. 2005, \aap, 430, 1059

\bibitem[{{Chandrasekhar}(1961)}]{Chandrasekhar61}
{Chandrasekhar}, S. 1961, {Hydrodynamic and hydromagnetic stability}

\bibitem[{{Churchwell} {et~al.}(2006){Churchwell}, {Povich}, {Allen}, {Taylor},
  {Meade}, {Babler}, {Indebetouw}, {Watson}, {Whitney}, {Wolfire}, {Bania},
  {Benjamin}, {Clemens}, {Cohen}, {Cyganowski}, {Jackson}, {Kobulnicky},
  {Mathis}, {Mercer}, {Stolovy}, {Uzpen}, {Watson}, \& {Wolff}}]{Churchwell+06}
{Churchwell}, E., {et~al.} 2006, \apj, 649, 759

\bibitem[{{Churchwell} {et~al.}(2007){Churchwell}, {Watson}, {Povich},
  {Taylor}, {Babler}, {Meade}, {Benjamin}, {Indebetouw}, \&
  {Whitney}}]{Churchwell+07}
---. 2007, \apj, 670, 428

\bibitem[{{Dale} {et~al.}(2011){Dale}, {W{\"u}nsch}, {Smith}, {Whitworth}, \&
  {Palou{\v s}}}]{Dale+11}
{Dale}, J.~E., {W{\"u}nsch}, R., {Smith}, R.~J., {Whitworth}, A., \& {Palou{\v
  s}}, J. 2011, \mnras, 411, 2230

\bibitem[{{Dale} {et~al.}(2009){Dale}, {W{\"u}nsch}, {Whitworth}, \& {Palou{\v
  s}}}]{Dale+09}
{Dale}, J.~E., {W{\"u}nsch}, R., {Whitworth}, A., \& {Palou{\v s}}, J. 2009,
  \mnras, 398, 1537

\bibitem[{{D'Alessio} {et~al.}(1998){D'Alessio}, {Canto}, {Calvet}, \&
  {Lizano}}]{Dalessio+98}
{D'Alessio}, P., {Canto}, J., {Calvet}, N., \& {Lizano}, S. 1998, \apj, 500,
  411

\bibitem[{{Deharveng} {et~al.}(2005){Deharveng}, {Zavagno}, \&
  {Caplan}}]{Deharveng+05}
{Deharveng}, L., {Zavagno}, A., \& {Caplan}, J. 2005, \aap, 433, 565

\bibitem[{{Dubois} \& {Teyssier}(2008)}]{Dubois+08}
{Dubois}, Y., \& {Teyssier}, R. 2008, \aap, 477, 79

\bibitem[{{Durisen} {et~al.}(2007){Durisen}, {Boss}, {Mayer}, {Nelson},
  {Quinn}, \& {Rice}}]{Durisen+07}
{Durisen}, R.~H., {Boss}, A.~P., {Mayer}, L., {Nelson}, A.~F., {Quinn}, T., \&
  {Rice}, W.~K.~M. 2007, Protostars and Planets V, 607

\bibitem[{{Elmegreen}(1982)}]{Elmegreen82}
{Elmegreen}, B.~G. 1982, \apj, 253, 634

\bibitem[{{Elmegreen}(1987)}]{Elmegreen87}
---. 1987, \apj, 312, 626

\bibitem[{{Elmegreen}(1989)}]{Elmegreen89}
---. 1989, \apj, 340, 786

\bibitem[{{Elmegreen}(1998)}]{Elmegreen98}
{Elmegreen}, B.~G. 1998, in Astronomical Society of the Pacific Conference
  Series, Vol. 148, Origins, ed. {C.~E.~Woodward, J.~M.~Shull, \&
  H.~A.~Thronson Jr.}, 150

\bibitem[{{Elmegreen} \& {Elmegreen}(1978)}]{Elmegreen+78}
{Elmegreen}, B.~G., \& {Elmegreen}, D.~M. 1978, \apj, 220, 1051

\bibitem[{{Elmegreen} \& {Lada}(1977)}]{Elmegreen+77}
{Elmegreen}, B.~G., \& {Lada}, C.~J. 1977, \apj, 214, 725

\bibitem[{{Goldreich} \& {Lynden-Bell}(1965{\natexlab{a}})}]{Goldreich+65a}
{Goldreich}, P., \& {Lynden-Bell}, D. 1965{\natexlab{a}}, \mnras, 130, 97

\bibitem[{{Goldreich} \& {Lynden-Bell}(1965{\natexlab{b}})}]{Goldreich+65b}
---. 1965{\natexlab{b}}, \mnras, 130, 125

\bibitem[{{Goldreich} \& {Tremaine}(1978)}]{Goldreich+78}
{Goldreich}, P., \& {Tremaine}, S. 1978, \apj, 222, 850

\bibitem[{{Harrison} \& {Lake}(1972)}]{Harrison+72}
{Harrison}, E.~R., \& {Lake}, R.~G. 1972, \apj, 171, 323

\bibitem[{{Hosokawa} \& {Inutsuka}(2006)}]{Hosokawa+06}
{Hosokawa}, T., \& {Inutsuka}, S.-i. 2006, \apj, 646, 240

\bibitem[{{Iwasaki} {et~al.}(2011){Iwasaki}, {Inutsuka}, \&
  {Tsuribe}}]{Iwasaki+11}
{Iwasaki}, K., {Inutsuka}, S.-i., \& {Tsuribe}, T. 2011, \apj, 733, 16

\bibitem[{{Julian} \& {Toomre}(1966)}]{Julian+66}
{Julian}, W.~H., \& {Toomre}, A. 1966, \apj, 146, 810

\bibitem[{{Kim} {et~al.}(2002){Kim}, {Ostriker}, \& {Stone}}]{Kim+02}
{Kim}, W.-T., {Ostriker}, E.~C., \& {Stone}, J.~M. 2002, \apj, 581,
1080

\bibitem[{{Kim} \& {Ostriker}(2007)}]{Kim+07}
{Kim}, W.-T., \& {Ostriker}, E.~C. 2007, \apj, 660, 1232

\bibitem[{{Kim} {et~al.}(2003){Kim}, {Ostriker}, \& {Stone}}]{Kim+03}
{Kim}, W.-T., {Ostriker}, E.~C., \& {Stone}, J.~M. 2003, \apj, 599, 1157

\bibitem[{{Korycansky} \& {Pringle}(1995)}]{Korycansky+95}
{Korycansky}, D.~G., \& {Pringle}, J.~E. 1995, \mnras, 272, 618

\bibitem[{{La Dous}(1994)}]{LaDous94}
{La Dous}, C. 1994, \ssr, 67, 1

\bibitem[{{Larson}(1985)}]{Larson85}
{Larson}, R.~B. 1985, \mnras, 214, 379

\bibitem[{{Latter} \& {Balbus}(2009)}]{Latter+09}
{Latter}, H.~N., \& {Balbus}, S.~A. 2009, \mnras, 399, 1058

\bibitem[{{Ledoux}(1951)}]{Ledoux51}
{Ledoux}, P. 1951, Annales d'Astrophysique, 14, 438

\bibitem[{{Lee} \& {Hong}(2007)}]{Lee+07}
{Lee}, S.~M., \& {Hong}, S.~S. 2007, \apjs, 169, 269

\bibitem[{{Lighthill}(1978)}]{Lighthill78}
{Lighthill}, J. 1978, {Waves in fluids} (Cambridge University Press)

\bibitem[{{Lin} {et~al.}(1990){Lin}, {Papaloizou}, \& {Savonije}}]{Lin+90}
{Lin}, D.~N.~C., {Papaloizou}, J.~C.~B., \& {Savonije}, G.~J. 1990, \apj, 364,
  326

\bibitem[{{Lubow} \& {Ogilvie}(1998)}]{Lubow+98}
{Lubow}, S.~H., \& {Ogilvie}, G.~I. 1998, \apj, 504, 983

\bibitem[{{Lubow} \& {Pringle}(1993)}]{Lubow+93}
{Lubow}, S.~H., \& {Pringle}, J.~E. 1993, \mnras, 263, 701

\bibitem[{{Mamatsashvili} \& {Rice}(2010)}]{Mamatsashvili+10}
{Mamatsashvili}, G.~R., \& {Rice}, W.~K.~M. 2010, \mnras, 406, 2050

\bibitem[{{McKee} \& {Ostriker}(2007)}]{McKee+07}
{McKee}, C.~F., \& {Ostriker}, E.~C. 2007, \araa, 45, 565

\bibitem[{{Mej{\'{\i}}a} {et~al.}(2005){Mej{\'{\i}}a}, {Durisen}, {Pickett}, \&
  {Cai}}]{Mejia+05}
{Mej{\'{\i}}a}, A.~C., {Durisen}, R.~H., {Pickett}, M.~K., \& {Cai}, K. 2005,
  \apj, 619, 1098

\bibitem[{{Nagai} {et~al.}(1998){Nagai}, {Inutsuka}, \& {Miyama}}]{Nagai+98}
{Nagai}, T., {Inutsuka}, S.-I., \& {Miyama}, S.~M. 1998, \apj, 506, 306

\bibitem[{{Nelson} {et~al.}(2000){Nelson}, {Benz}, \& {Ruzmaikina}}]{Nelson+98}
{Nelson}, A.~F., {Benz}, W., \& {Ruzmaikina}, T.~V. 2000, \apj, 529, 357

\bibitem[{{Ogilvie} \& {Lubow}(1999)}]{Ogilvie+99}
{Ogilvie}, G.~I., \& {Lubow}, S.~H. 1999, \apj, 515, 767

\bibitem[{{Rice} {et~al.}(2003){Rice}, {Armitage}, {Bate}, \&
  {Bonnell}}]{Rice+03}
{Rice}, W.~K.~M., {Armitage}, P.~J., {Bate}, M.~R., \& {Bonnell}, I.~A. 2003,
  \mnras, 339, 1025

\bibitem[{{Rice} {et~al.}(2005){Rice}, {Lodato}, \& {Armitage}}]{Rice+05}
{Rice}, W.~K.~M., {Lodato}, G., \& {Armitage}, P.~J. 2005, \mnras, 364, L56

\bibitem[{{Safronov}(1960)}]{Safronov60}
{Safronov}, V.~S. 1960, Annales d'Astrophysique, 23, 979

\bibitem[{{Simon}(1965)}]{Simon65}
{Simon}, R. 1965, Annales d'Astrophysique, 28, 40

\bibitem[{{Toomre}(1964)}]{Toomre64}
{Toomre}, A. 1964, \apj, 139, 1217

\bibitem[{{Umekawa} {et~al.}(1999){Umekawa}, {Matsumoto}, {Miyaji}, \&
  {Yoshida}}]{Umekawa+99}
{Umekawa}, M., {Matsumoto}, R., {Miyaji}, S., \& {Yoshida}, T. 1999, \pasj, 51,
  625

\bibitem[{{Usami} {et~al.}(1995){Usami}, {Hanawa}, \& {Fujimoto}}]{Usami+95}
{Usami}, M., {Hanawa}, T., \& {Fujimoto}, M. 1995, \pasj, 47, 271

\bibitem[{{Viala} \& {Horedt}(1974)}]{Viala+74a}
{Viala}, Y., \& {Horedt}, G.~P. 1974, \aaps, 16, 173

\bibitem[{{Voit}(1988)}]{Voit88}
{Voit}, G.~M. 1988, \apj, 331, 343

\bibitem[{{Whittaker} \& {Watson}(1963)}]{Whittaker+63}
{Whittaker}, E.~T., \& {Watson}, G.~N. 1963, {A course of modern
analysis}

\bibitem[{{W{\"u}nsch} {et~al.}(2010){W{\"u}nsch}, {Dale}, {Palou{\v s}}, \&
  {Whitworth}}]{Wunsch+10}
{W{\"u}nsch}, R., {Dale}, J.~E., {Palou{\v s}}, J., \& {Whitworth}, A.~P. 2010,
  \mnras, 407, 1963

\bibitem[{{W{\"u}nsch} {et~al.}(2012){W{\"u}nsch}, {J{\'a}chym}, {Sidorin},
  {Ehlerov{\'a}}, {Palou{\v s}}, {Dale}, {Dawson}, \& {Fukui}}]{Wunsch+12}
{W{\"u}nsch}, R., {J{\'a}chym}, P., {Sidorin}, V., {Ehlerov{\'a}}, S.,
  {Palou{\v s}}, J., {Dale}, J., {Dawson}, J.~R., \& {Fukui}, Y. 2012, \aap,
  539, A116

\bibitem[{{Zapatero Osorio} {et~al.}(2002){Zapatero Osorio}, {B{\'e}jar},
  {Mart{\'{\i}}n}, {Rebolo}, {Barrado y Navascu{\'e}s}, {Mundt},
  {Eisl{\"o}ffel}, \& {Caballero}}]{Zapatero+02}
{Zapatero Osorio}, M.~R., {B{\'e}jar}, V.~J.~S., {Mart{\'{\i}}n}, E.~L.,
  {Rebolo}, R., {Barrado y Navascu{\'e}s}, D., {Mundt}, R., {Eisl{\"o}ffel},
  J., \& {Caballero}, J.~A. 2002, \apj, 578, 536

\bibitem[{{Zel'Dovich}(1970)}]{Zeldovich70}
{Zel'Dovich}, Y.~B. 1970, \aap, 5, 84

\end{thebibliography}
\end{document}